\documentclass[fleqn,usenatbib,twocolumn]{mnras}
\usepackage{newtxtext,newtxmath,subfigure}
\usepackage[T1]{fontenc}

\usepackage{booktabs}
\usepackage{lscape}
\usepackage{graphicx}	% Including figure files
\usepackage{amsmath}	% Advanced maths commands
\usepackage{multicol}
\usepackage{caption}
\usepackage{placeins}
\usepackage{rotating,longtable}

\newcommand{\lum}{erg s$^{\rm -1}$\,}
\newcommand{\density}{cm$^{\rm -3}$\,}
\newcommand{\vol}{cm$^{\rm 3}$\,}
\newcommand{\pressure}{dyne cm$^{\rm -2}$\,}

\newcommand{\lxf}{ L$_{\rm XF}$\,}
\newcommand{\lxq}{ L$_{\rm XQ}$\,}
\newcommand{\nh}{{\rm $N_H$}\,}
\title[X-ray flares from AB Dor]{Study of the Energetic X-ray Superflares from the active fast rotator AB Doradus}

\author[Shweta Didel et al.]{
Shweta Didel$^{1}$\thanks{E-mail: shwetadidel.rs.phy19@itbhu.ac.in}, 
Jeewan C. Pandey$^{2}$\thanks{E-mail: jeewan.aries@nic.in},
A.K.~Srivastava$^{1}$\thanks{E-mail: asrivastava.app@itbhu.ac.in},
 and Gurpreet Singh$^{2}$
\\
$^{1}$Department of Physics, Indian Institute of Technology (BHU), Varanasi-221005, India \\
$^{2}$Aryabhatta Research Institute of Observational Sciences (ARIES), Manora Peak, Nainital-2632001, India \\
}

\begin{document}
\label{firstpage}
\pagerange{\pageref{firstpage}--\pageref{lastpage}}
\maketitle

\begin{abstract}
We present the analyses of intense X-ray flares detected on the active fast rotator AB Dor using observations from the XMM-Newton. A total of 21 flares are detected, and 13 flares are analysed in detail. The total X-ray energy of these flares is found to be in the range of 10$^{34-36}$ erg, in which the peak flare flux increased up to 34 times from the pre-/post-flaring states for the strongest observed flare. The duration of these flaring events is found to be 0.7 to 5.8 hrs. The quiescent state X-ray spectra are found to be explained by a three-temperature plasma with average temperatures of 0.29, 0.95, and 1.9 keV, respectively. The temperatures, emission measures, and abundances are found to be varying during the flares. The peak flare temperature was found in the 31-89 MK range, whereas the peak emission measure was 10$^{52.5-54.7}$ \density. The abundances vary during the flares and increase by a factor of $\sim$3 from the quiescent value for the strongest detected flare. The variation in individual abundances follows the inverse-FIP effect in quiescent and flare phases. The X-ray light curves of AB Dor are found to exhibit rotational modulation. The semi-loop lengths of the flaring events are derived in the range of 10$^{9.9-10.7}$ cm, whereas the minimum magnetic field to confine the plasma in the flaring loop is estimated between 200 and 700 G.

\end{abstract}


\section{Introduction}
\label{sec:intro}
The X-ray emissions from the late-type stars are well explained by the magnetic confinement of million-degree plasma residing in the coronal loops \citep[see][and references therein]{2014LRSP...11....4R}. The magnetic field is supposed to be generated by the dynamo process working deep beneath the convective shells of the late-type stars. The presence of strong magnetic activities is indicated by the fact that strong X-ray emissions are observed even in their quiescent phase by the fast-rotating late-type stars, and the star spot covers large fractions of the stellar surface.
Stellar coronae of these stars show a very dynamic nature of various activities, e.g., short explosive bursts of energy in time from a few minutes to several hours and slowly varying quiescent corona over a year to a few years. These short explosive bursts on stars are commonly known as flares. The magnetic reconnection process explains the energy released during the flares, which ranges from $10^{30-38}$ \lum \citep[][]{1988ApJ...330..474P,2010ARA&A..48..241B}.
 Since the stellar coronae are nearly 1000 times brighter than the Sun, the flares on stars are more energetic and impulsive than the Sun. The energetic flares having energy release $\ge10^{33}$ \lum are termed as superflares \citep[][]{2000ApJ...529.1026S, 2013ApJS..209....5S, 2016PASJ...68...90T, 2022A&A...666A.198P}.

\begin{table*}
	\centering
	\caption{Log of observations of AB Dor with XMM-Newton. }
	\label{tab:log_table}
	\begin{tabular}{cclccccccc} 
		\hline\hline
		Set & Observation & Instrument & Start time & Exposure & Offset & Src Radius & Bkg Radius & Mode & Filter\\
		    & ID & & UTC & (s) & ($^\prime$) & (") & (") &\\
		\hline
		 S1 & 0123720301 & PN & 27/10/2000 \enspace15:23:55 & 55700 & 0.145 & 70 & 50,50 & Small Window & Medium\\ 
		    & & MOS & 27/10/2000 \enspace15:10:06 & 56049 & 0.145 & 110 & 110 & Full Frame & Medium\\ 
		    & & RGS & 27/10/2000  \enspace15:01:40 &  58908 &  0.145 &&\\ 
		 S2 & 0134520701 & PN   & 22/05/2001 \enspace 17:05:58 & 48220 & 0.145 & 70 & 50,50 & Small Window & Medium\\
		    & & MOS & 22/05/2001 \enspace16:50:15 & 49019 & 0.145 & 80 & 80 & Full Frame & Medium\\
		    & & RGS  & 22/05/2001 \enspace16:43:55 & 49607 & 0.145 &&\\
		 S3 & 0412580701 & PN & 03/01/2011 \enspace02:10:27 & 10000 & 0.009 & 7.5,55* & 45,32 & Small Window & Thick\\
		    & & MOS & 03/01/2011 \enspace02:07:20 & 12300 & 0.009 & 50 & 50 & Small Window & Thick\\
		    & & RGS  & 02/01/2011 \enspace16:58:57 & 56051 & 0.009 &&\\
		 S4 & 0412580801 & PN & 01/01/2012 \enspace02:09:42 & 10001 & 0.009 & 65 & 46,46 & Small Window & Thick\\
		    & & MOS & 01/01/2012 \enspace02:06:59 & 9999 & 0.009 & 48 & 48 & Small Window & Thick\\
		    & & RGS  & 31/12/2011 \enspace15:41:48 & 60770 & 0.009 &&\\
		 S5 & 0791980101 & PN & 07/10/2016 \enspace01:32:25 & 10019 & 0.009 & 20,68* & 45,45,25 & Small Window & Thick\\ 
		    & & MOS & 07/10/2016 \enspace01:26:56 & 10077 & 0.009 & 54 & 54 & Small Window & Thick\\ 
		    & & RGS  & 07/10/2016 \enspace01:26:47 & 98313 & 0.009 &&\\ 
		 S6 & 0810850501 & PN   & 30/09/2019 \enspace22:26:59 & 12000 & 0.009 & 60 & 45,40 & Small Window & Thick\\
		    & & MOS & 30/09/2019 \enspace22:18:37 & 11999 & 0.009 & 47 & 47 & Small Window & Thick\\
		    & & RGS  & 30/09/2019 \enspace01:55:59 & 99485 & 0.009 &&\\

		\hline
	\end{tabular}
    \\
     ~~~* The selected region corresponds to the inner and outer radii of the annulus region used for pile-up removal.
\end{table*}

In the standard flares model, the magnetic reconnection process causes a rapid and transient catastrophic release of magnetic energy in the corona, leading to particle acceleration and electromagnetic radiation ranging from radio waves to $\gamma$-rays. These accelerated charged particles gyrate downward along magnetic field lines, resulting in synchrotron radio emission. Hard X-rays are produced when these electron and proton beams collide with denser chromospheric materials. Simultaneous heating of the plasma to tens of millions of Kelvin evaporates cold material from the chromospheric footpoints, thus increasing the density of newly formed coronal loops and emitting UV and soft X-rays. 

Stellar X-ray flares were extensively studied by several authors in the past \cite[e.g.][] {1990A&A...228..403P,2008ApJ...688..418G,2008MNRAS.387.1627P,2012MNRAS.419.1219P}. They come in various durations (ranging from a few minutes to a few hours), shapes, and sizes. \cite{1988A&A...205..181V} found flares with short rise time as compact ones, while flares having longer rise time must be the two-ribbon flares. 
In some of the giant flares, double exponential decays are reported \citep[see, e.g.][]{1999ApJ...515..746O,1999A&A...350..900F,2000A&A...353..987F}. 
Superflares could also give information about the extent of coronae, e.g., if the flare is visible during a full rotation of a star, it means either the flaring region is very extended \citep[][]{1996A&A...311..211K} or the flare occurred near the pole \citep[][]{2000A&A...356..627M}.  
Researchers have detected stellar superflares in optical and UV bands from a variety of sources, including RS CVn binaries, young T-Tauri stars, UV Ceti-like red dwarfs, and solar-like G-dwarfs \citep[e.g.][] {1991ARA&A..29..275H,2012Natur.485..478M,2021ApJ...912...81K}. However, X-ray superflares have been observed in a few stars so far, e.g., Algol \citep[][]{1999A&A...350..900F}, AB Dor \citep[][]{2000A&A...356..627M}, II Peg \citep[][]{2007ApJ...654.1052O}, CC Eri \citep[][]{2017ApJ...840..102K}, and SZ Psc \citep[][]{2023MNRAS.518..900K}.

Unlike Solar flares, stellar flares are spatially unresolved; however, a study of the complete evolution of a stellar ﬂare can allow us to access information about the plasma structure and morphology.
Coronal abundances are also found to be changing during the large flares \citep[e.g.][]{1989PASJ...41..679T,1992ApJ...400..321S,2012MNRAS.419.1219P,2017ApJ...840..102K}, which suggest that flaring topologies can alter the fractionation processes in corona. This paper presents a detailed analysis of intense X-ray flares observed in the active star AB Dor A.

AB Dor A is a young, active, and a member of a pre-main-sequence quintuplet stellar system AB Doradus (=HD 36705) at a distance of $15.0\pm0.1$ pc \citep[][]{1997ApJ...490..835G}. The other companions, binaries AB Dor Ca/Cb \citep[][]{2019ApJ...886L...9C} and AB Dor Ba/Bb \citep[][]{1987PASP...99.1071V} are faint sources and are ignored in the following work. Hereafter, we refer to AB Dor A as AB Dor throughout the text. AB Dor is a highly active ultra-fast rotator of K0V spectral type with a rotational period of 0.5148 days \citep[][]{1981A&A...104...33P}. It shows frequent flaring activity, probably due to its high spin rate ($\sim$50 times that of the Sun). Therefore, it possesses a strong magnetic field. It has a radius (R$_*$) of $\sim$0.96 R$_{\odot}$, a mass of 0.86 M$_{\odot}$, and a surface temperature of 5081 K\citep[see][and references therein]{2011A&A...533A.106G}.
In the X-ray band, AB Dor has gained interest since its first detection by Einstein Observatory \citep[][]{1981A&A...104...33P}. AB Dor is observed by almost every X-ray satellite because of its X-ray brightness and positional advantage (galactic latitude $\sim$ -33$^\circ$). \cite{1997A&A...320..831K} studied long-term X-ray activities using ROSAT data and found no significant trend throughout five years. Almost all X-ray observatories have detected frequent flares in the corona of AB Dor. \cite{1993A&A...278..467V} first reported statistical studies of X-ray flares where they found mean flare energy of $\sim10^{34}$ erg, putting AB Dor as a frequent super-flaring star.
In this context, we conducted a detailed study of X-ray flares on AB Dor using the XMM-Newton Observatory.

We organise the paper as follows. Section \ref{sec:obs} gives the observations and the principle data reduction procedures. In Section \ref{sec:results}, we have explained the analysis procedure in detail and showed the scientific results obtained from X-ray timing and spectral analysis. Finally, in Sections \ref{sec:discussion} and \ref{sec:summary}, we have discussed the results and presented our conclusion.

\begin{figure*}
\centering
%\vspace{-1.0cm}

\subfigure[Set 1]{\includegraphics[width=0.93\columnwidth,trim={1.0cm 1.0cm 0.0cm 3.5cm}, clip]{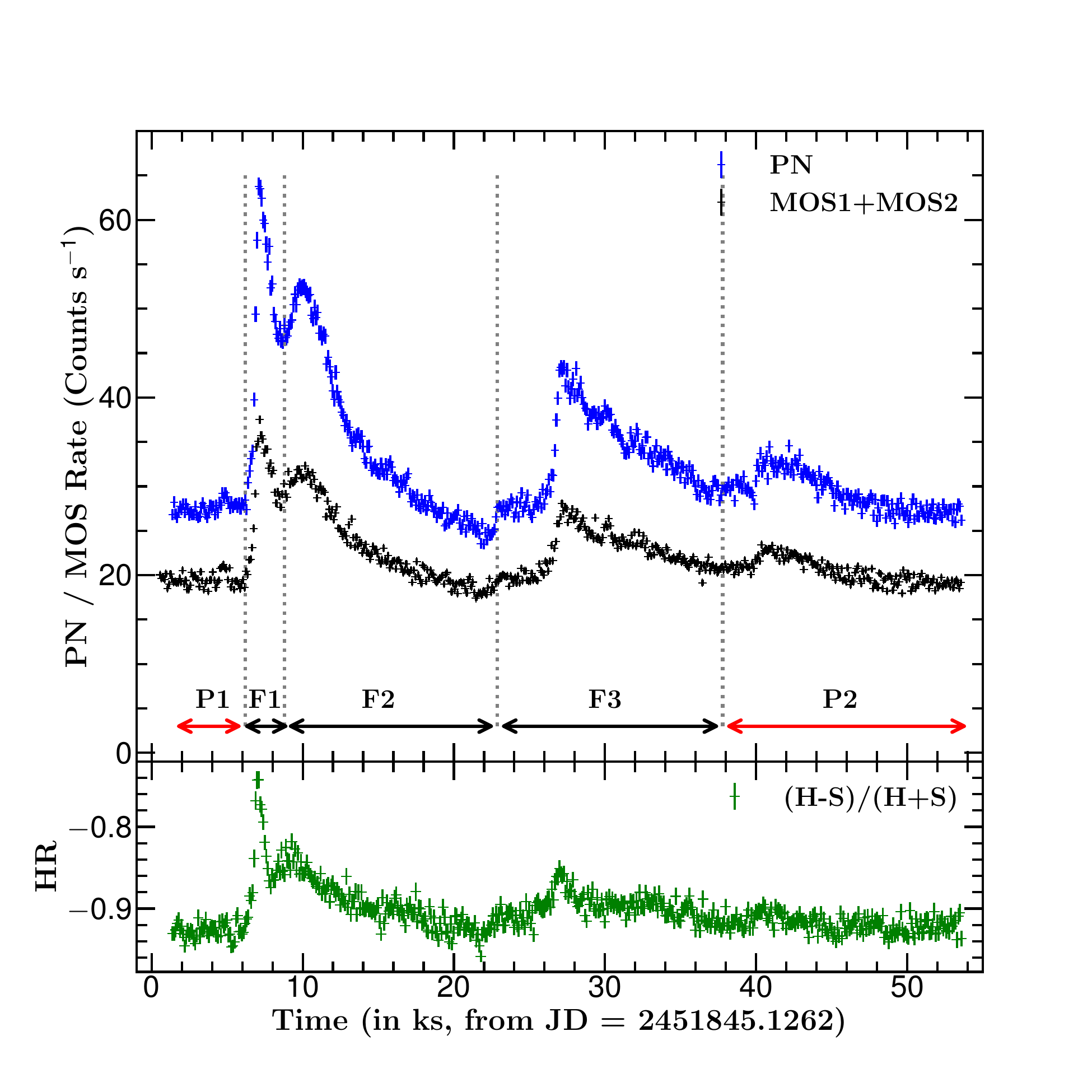}}
\subfigure[Set 2]{\includegraphics[width=0.93\columnwidth,trim={1.0cm 1.0cm 0.0cm 3.5cm}, clip]{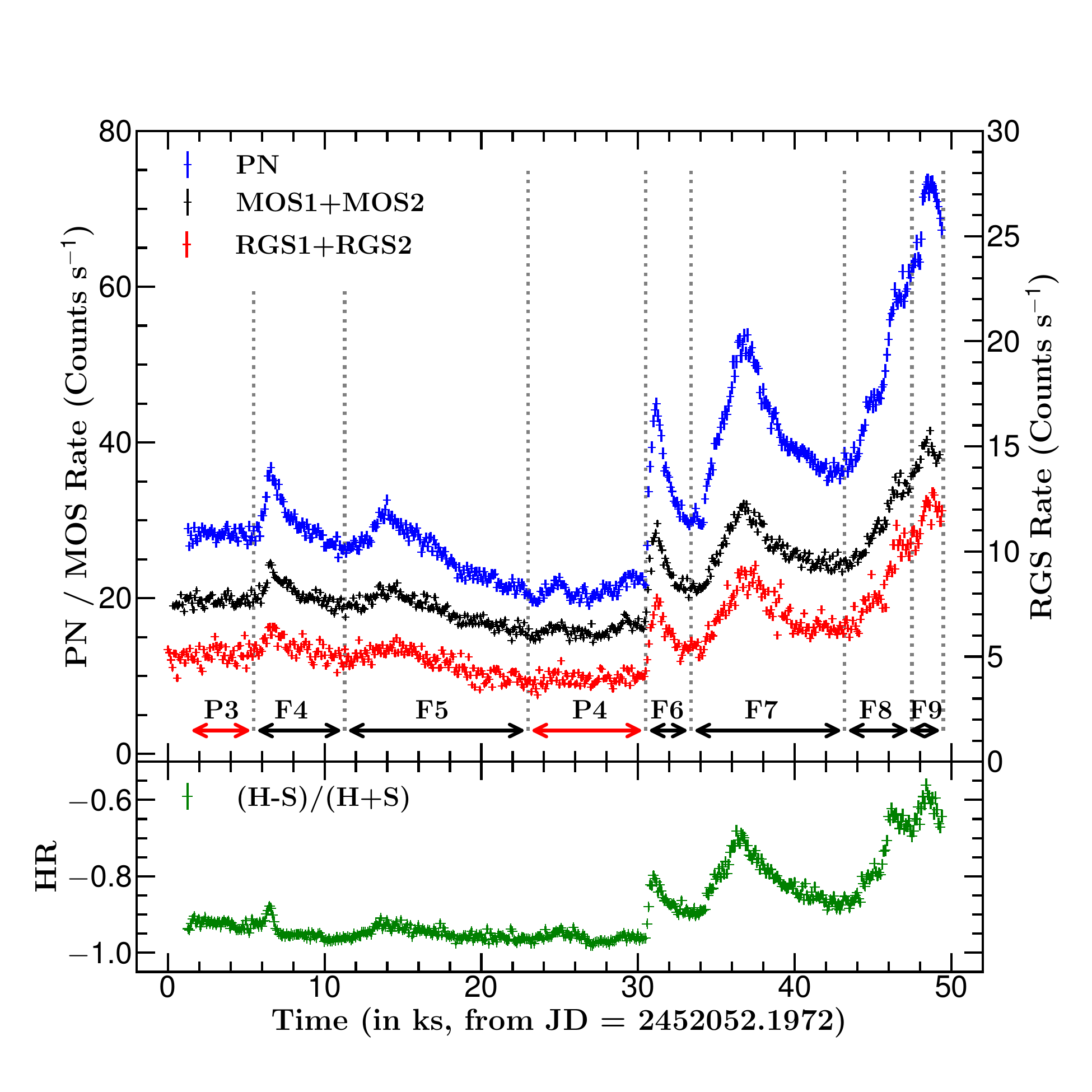}}
\vspace{-0.3cm}
\subfigure[Set 3]{\includegraphics[width=0.93\columnwidth,trim={1.0cm 1.0cm 0.0cm 3.5cm}, clip]{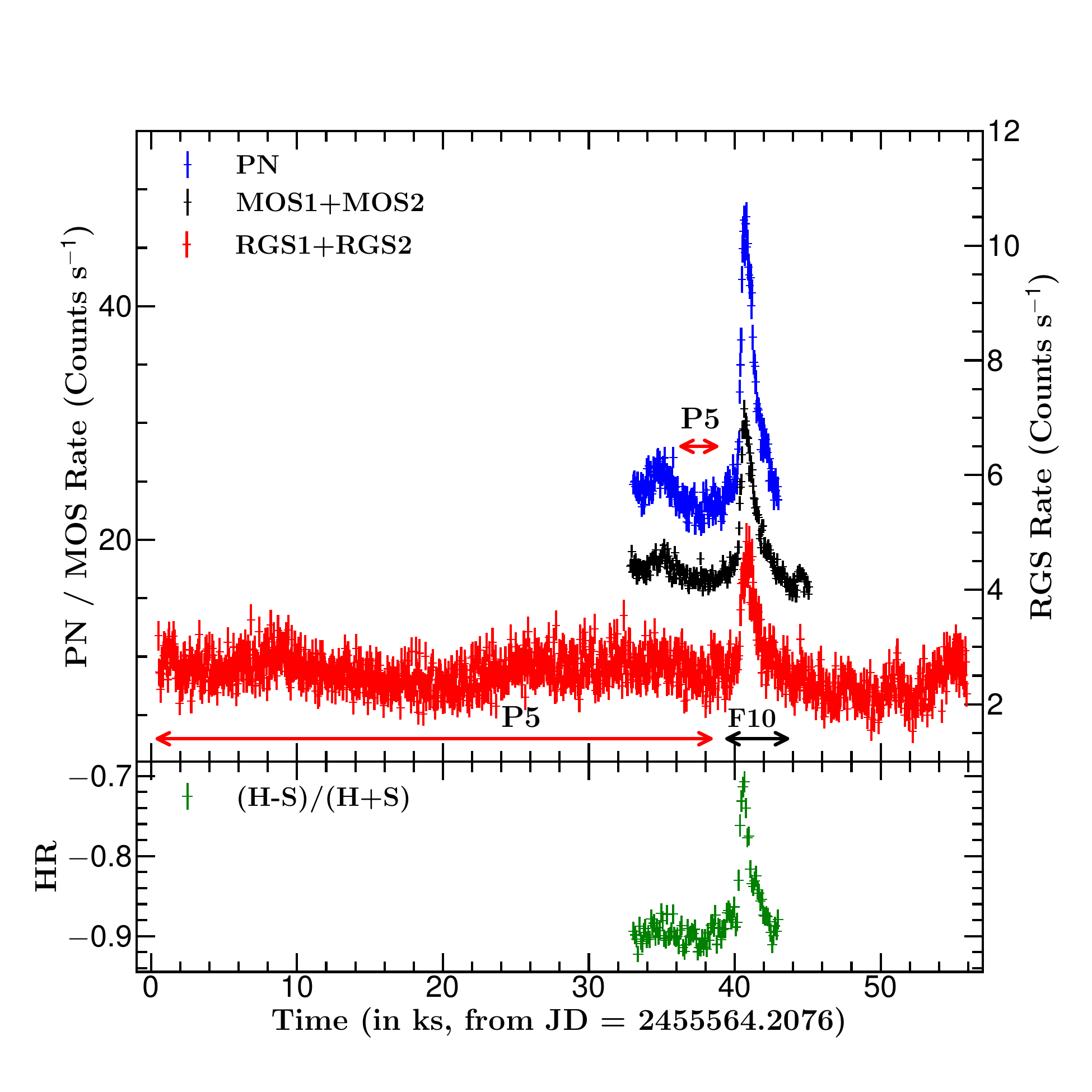}}
\subfigure[Set 4]{\includegraphics[width=0.93\columnwidth,trim={1.0cm 1.0cm 0.0cm 3.5cm}, clip]{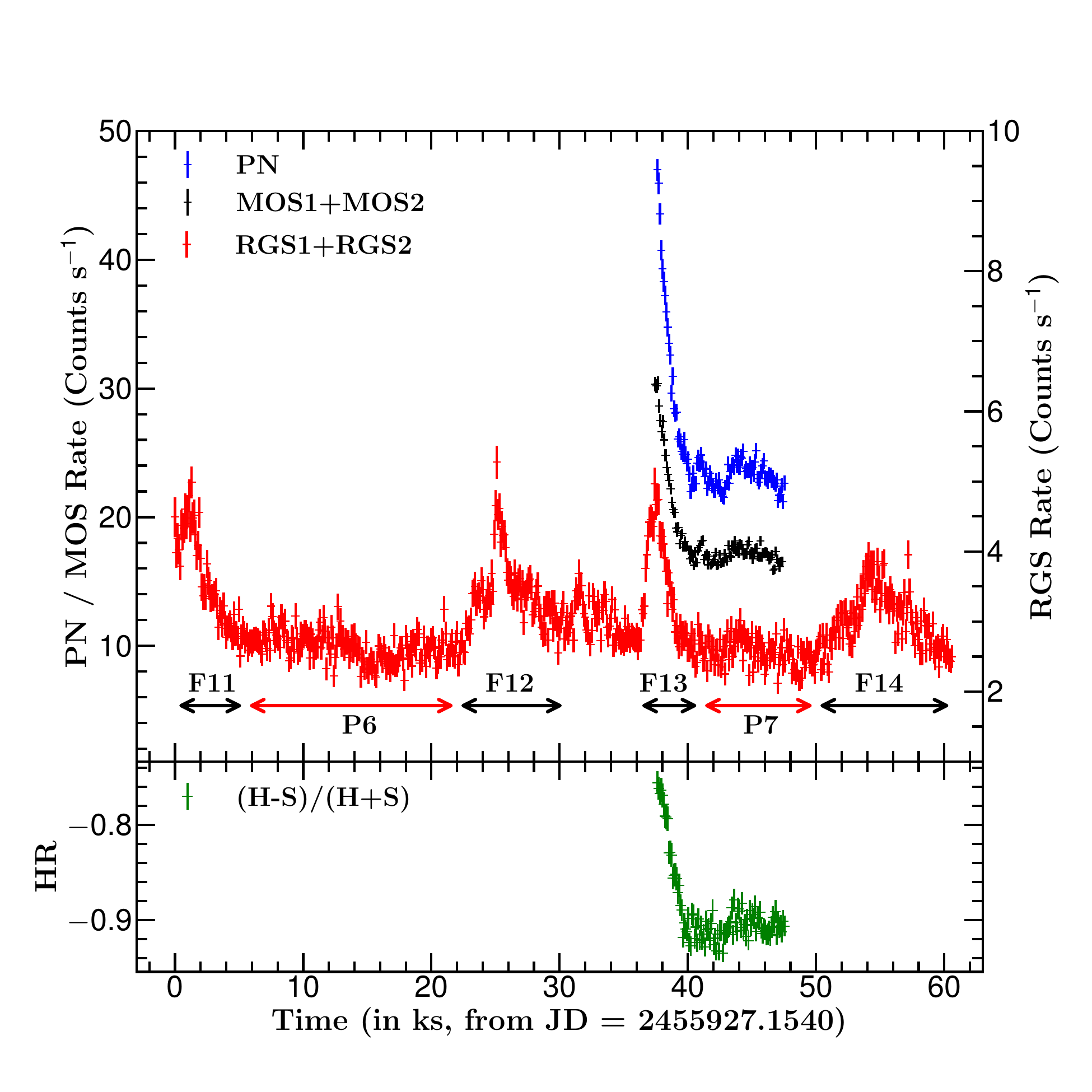}}
\vspace{-0.3cm}
\subfigure[Set 5]{\includegraphics[width=0.93\columnwidth,trim={1.0cm 1.0cm 0.0cm 3.5cm}, clip]{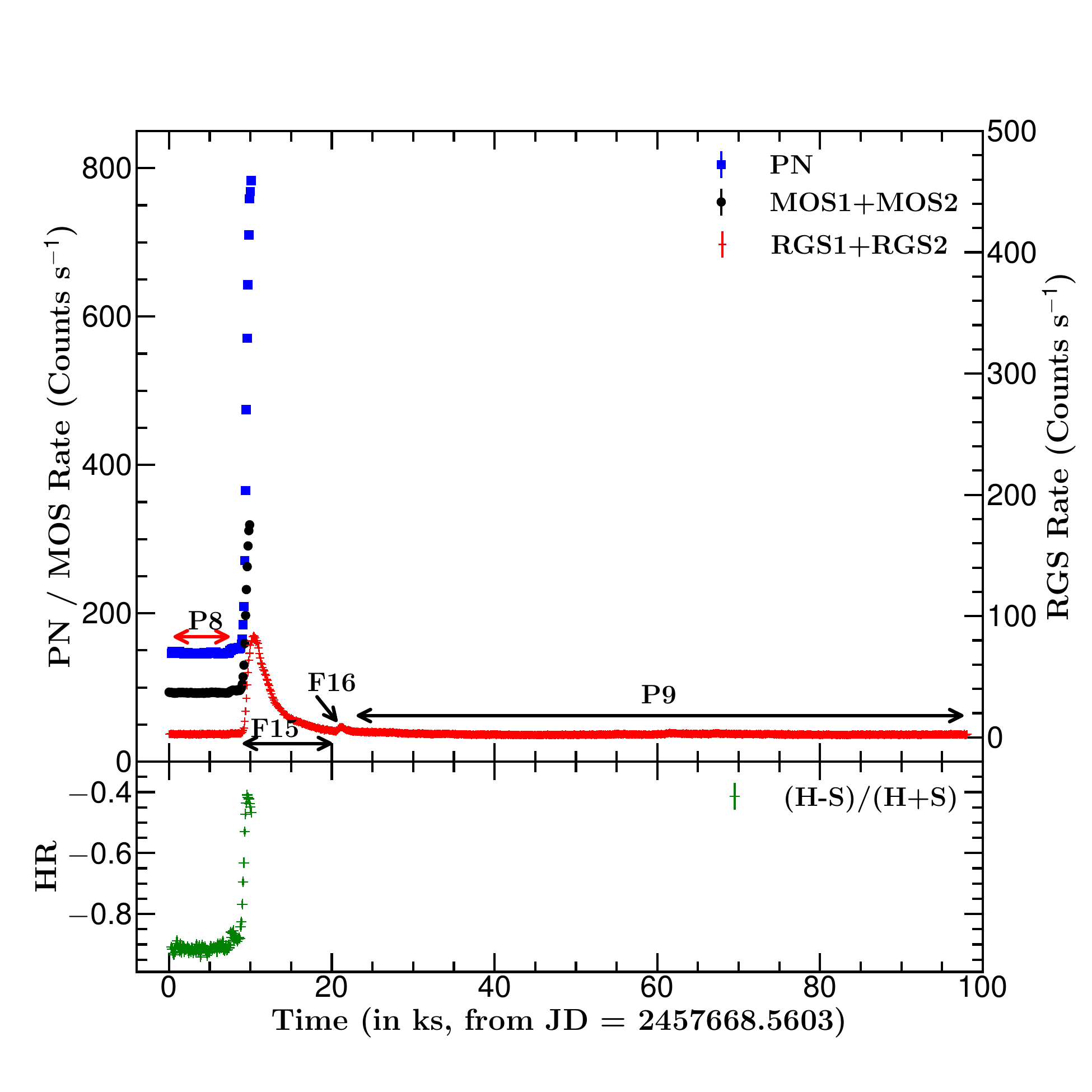}}
\subfigure[Set 6]{\includegraphics[width=0.93\columnwidth,trim={1.0cm 1.0cm 0.0cm 3.5cm}, clip]{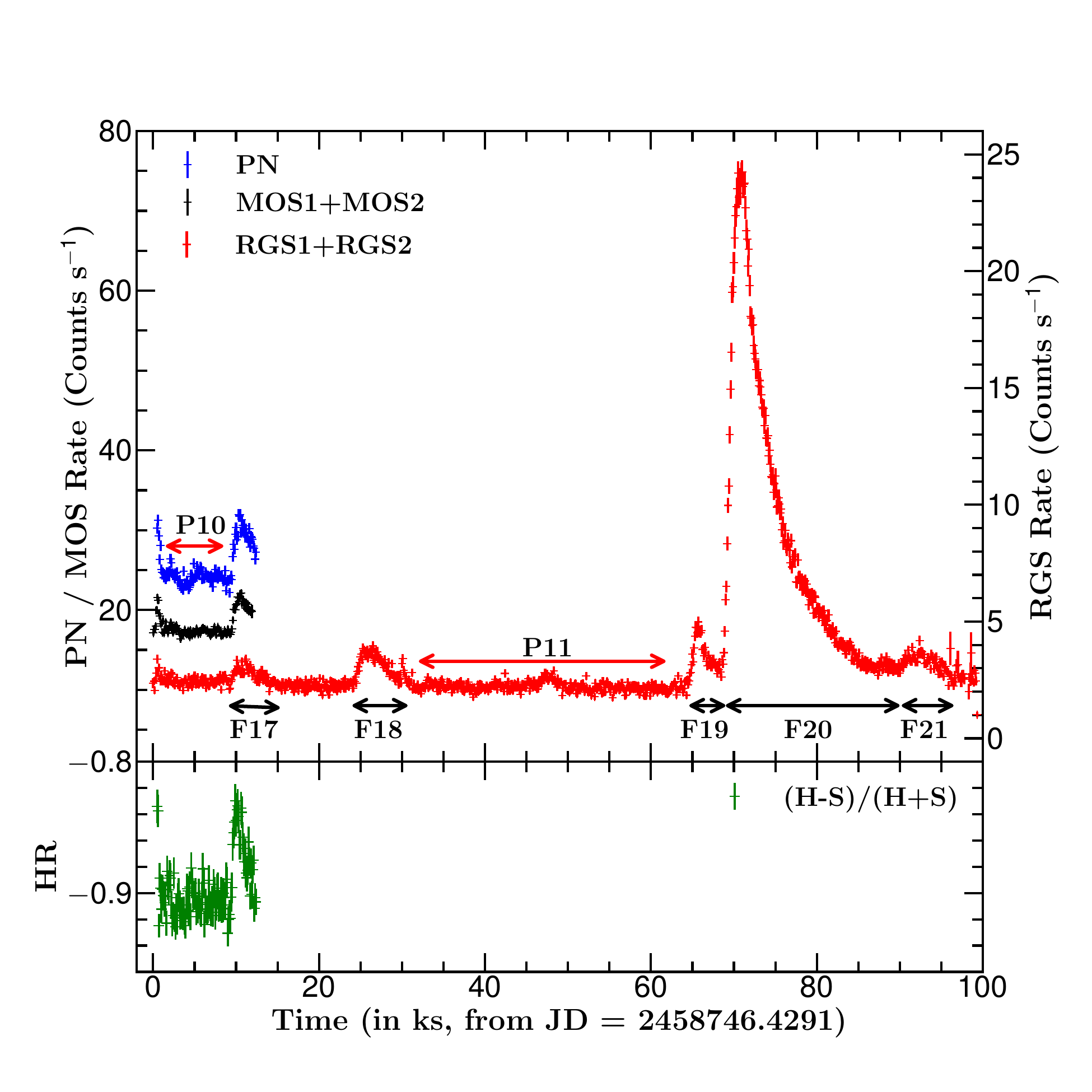}}
\vspace{-0.1cm}
\caption{The background subtracted AB Dor's X-ray light curves for different observation epochs. Light curves from PN, MOS, and RGS detectors are shown in blue, black, and red colours, respectively. In order to plot the light curves from all three detectors in the same panel, the MOS count rates are scaled up as 1.8 $\times$ MOS for sets S1 and S2, 5+MOS for sets S3, S4, and S6, and 80+MOS for set S5. PN count rate is scaled up as 120+PN for set S5 and RGS count rate is scaled up as 1.8 $\times$ RGS for set S2.}
\label{fig:lcs_all_HR}
\end{figure*}
\section{Observations and Data Reduction}
\label{sec:obs} 
AB Dor A was observed using the Reflection Grating Spectrometers (RGS ; \citep[][]{2001A&A...365L...7D}) and European Photon Imaging Cameras (EPIC; MOS \citep[][]{2001A&A...365L..27T} and PN \citep[][]{2001A&A...365L..18S}) aboard the well known XMM-Newton \citep[][]{2001A&A...365L...1J}. 
The EPIC cameras provide moderate spectral resolution (20-50 E/$\Delta$E) and good angular resolution (6" PSF-FWHM) in the energy range of 0.1 - 15 keV. The RGS instruments offer substantially better spectral resolution (150 -- 800) in the energy range of 0.33 - 2.5 keV (or 5 - 35 \AA).
The log of X-ray observations analysed in this paper is given in Table \ref{tab:log_table}.

The Science Analysis System (SAS) software version 18.0.0 of XMM-Newton and the updated calibration files were used to perform the EPIC and RGS data reduction. The raw EPIC data was processed to generate the event files using the tasks \textsc{epproc} and \textsc{emproc}, respectively, for PN and MOS data. Due to the high background contribution at high energies, we have selected the energy range between 0.3 to 10.0 keV to analyse EPIC data. Further, to identify the high background proton flaring intervals from the event file, we used the task \textsc{evselect} for energy greater than 10 keV. All the data sets are found to be free from such intense proton flaring events. We also examined the pile-up effect data using the task {\sc epatplot}. It was significant for the observation of set S5, and a very little pile-up was present in the case of set S3. This pile-up effect was limited to the observation from PN CCD only.
X-ray spectra and light curves from all EPIC observations were extracted from on-source counts obtained from circular regions around the source, whereas the background was chosen from source-free regions near the source in the same CCD. 
In order to avoid the pile-up effect for the observations S3 and S5, we chose annulus regions with inner and outer radii of 7.5" and 55", and 20" and 68", respectively. 
All the X-ray light curves obtained from EPIC observations were corrected for background contribution and other effects using task {\sc epiclccorr}. Light curves from MOS1 and MOS2 detectors were added using \textsc{lcmath} task.  We used the \textsc{especget} task to create source and background spectra with redistribution matrix (RMF) and auxiliary (ARF) files. We then binned all the X-ray spectra to have a minimum of 20 counts per bin using the \textsc{grppha} task.

The raw RGS data was reduced using the task \textsc{rgsproc} to generate event files and other spectral products; however, tasks \textsc{rgslccorr} and \textsc{rgcombine} were used to create RGS1 and RGS2 combined light curve and spectra, respectively.
These grouped spectra were used for further analysis.

\section{Analysis and Results}
\label{sec:results}
\subsection{X-ray light curves}
\label{sec:FLC}
The upper panels of Figure \ref{fig:lcs_all_HR} show the background-subtracted X-ray light curves of AB Dor as obtained from EPIC and RGS instruments in the energy range of 0.3 -- 10 keV and 0.33 -- 2.07 keV, respectively, whereas the lower panels show the variation of hardness ratio (HR) with time, which is defined as (H-S)/(H+S), where H is the count rate in the hard energy band of 2.0 -- 10.0 keV and S is the count rate in the soft energy band of 0.3 -- 2.0 keV. 

\begin{figure*}
\centering
\subfigure[set S2]{\includegraphics[width=0.93\columnwidth,trim={1.0cm 1.3cm 3.0cm 3.5cm},clip]{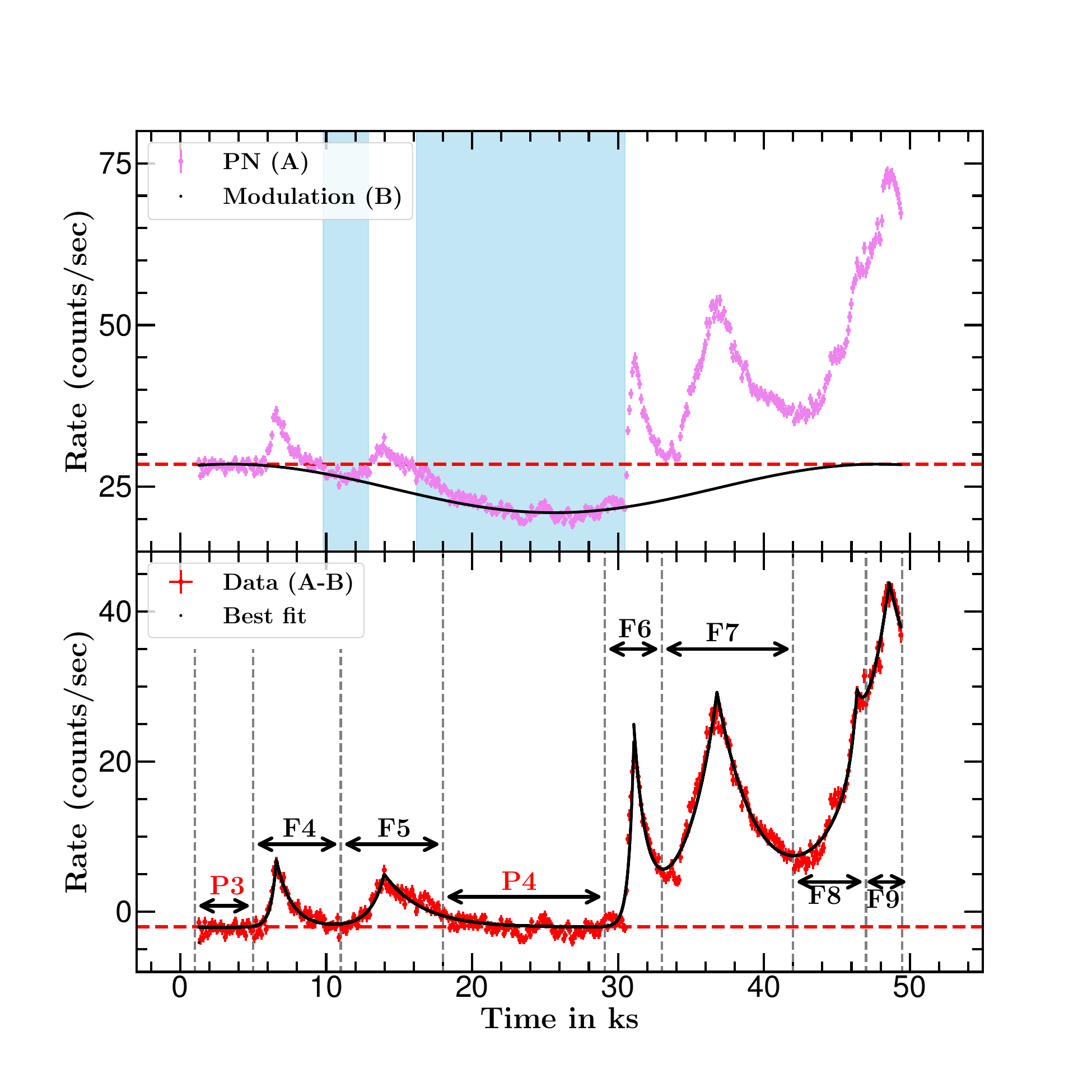}}
\subfigure[set S3]{\includegraphics[width=0.93\columnwidth,trim={1.0cm 1.3cm 3.0cm 3.5cm},clip]{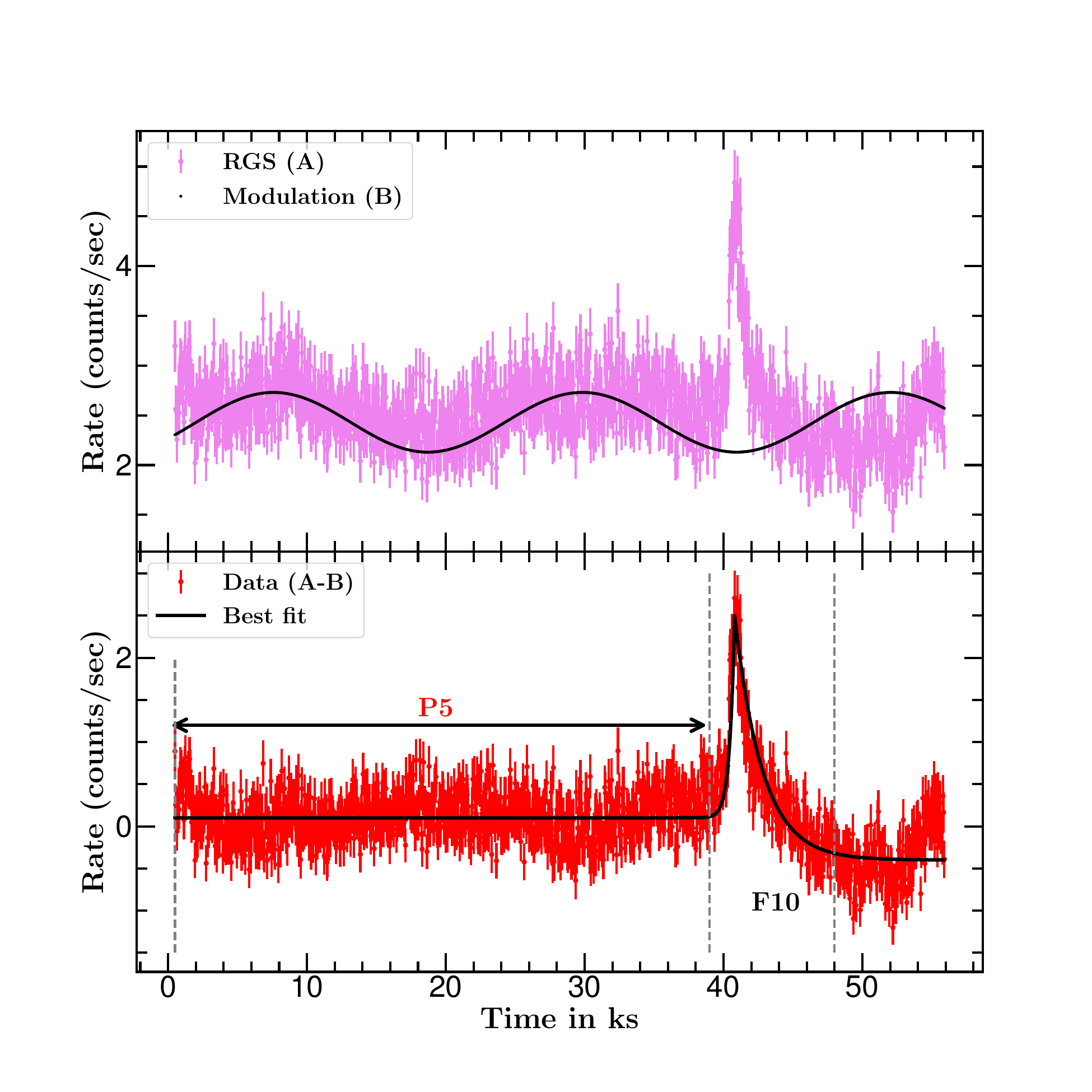}}
%\vspace{-0.3cm}
\subfigure[set S4]{\includegraphics[width=0.93\columnwidth,trim={0.7cm 1.0cm 2.0cm 3.5cm},clip]{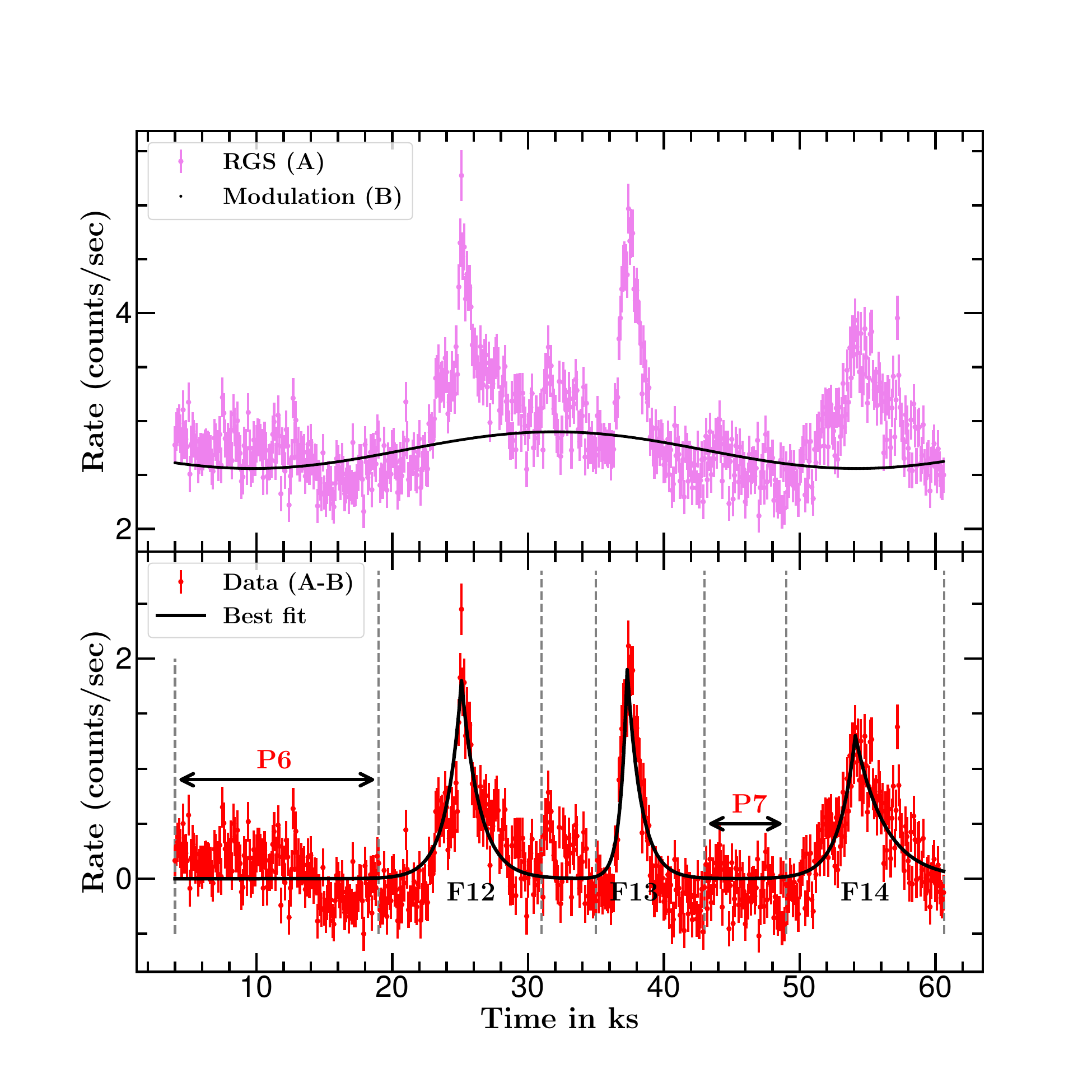}}
\subfigure[set S5]{\includegraphics[width=0.93\columnwidth,trim={0.7cm 1.0cm 2.0cm 3.5cm},clip]{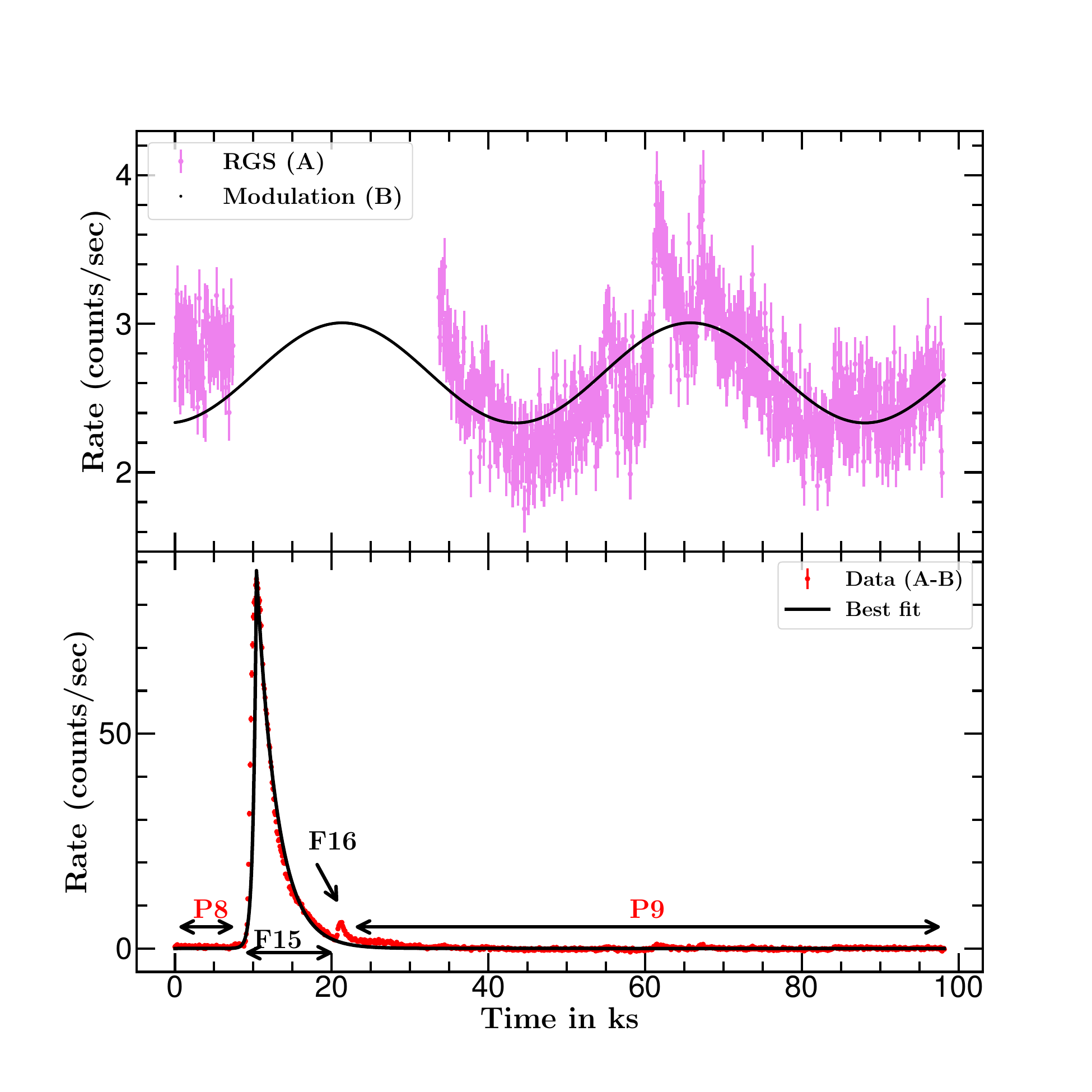}}

\caption{In the upper panels, PN/RGS light curve (A) and the presence of rotational modulation with best-fit sin curve (B) are shown in violet and black, respectively, for sets S2, S3, S4, and S5. To show the rotational modulation in set S5, the flare part is not shown due to the very high count rate during this flare F15. The best-fit flare model on the residual (A-B) is shown in the lower panel in black. The regions shaded in light blue colour in the upper panel of (a) represent the dimming areas as defined by \citep{2021NatAs...5..697V}, and the red dotted lines show the level of pre-flare average counts.}

\label{fig:rot_mod}
\end{figure*}

The X-ray light curves of AB Dor also exhibit rotational modulations \citep[e.g.,][]{1988MNRAS.231..131C,1993A&A...278..467V,1997A&A...320..831K}. Therefore, we used HR as a proxy to detect flaring events. We considered excursions greater than three times the standard deviation in the positive count rate from the pre/post-flare light curves to identify flares. Flaring regions in the light curves were identified when HR increased and mimicked the flare light curve. Using this approach, we detected 21 flares during the observations used here. We marked all the identified flare durations as Fi (where i = 1, 2, ... 21) in Figure \ref{fig:lcs_all_HR}. Regions of constant HR beyond the flaring region of the light curves were identified as either pre or post-flare epochs and were marked by Pi (where i = 1, 2, ... 11) in the same figure. During the pre-/post-flare segments, HR remained almost constant.
We fitted the following equation to model the light curve: the pre/post-flare segments were fitted with a horizontal straight line, and the rise and decay phases of the flares were fitted with an exponential function.

\begin{equation}
    c(t) = \begin{cases} c_{0}, & \mbox{for } t\leq t_{0} \\ 
    c_{0}\exp(\frac{t-t_{p}}{\tau_{r}}), & \mbox{for } t_{0}\leq t \leq t_{p} \\ 
    c_{0}\exp({-\frac{t-t_{p}}{\tau_{d}}}), & \mbox{for } t \geq t_{p} \\
\end{cases}
\end{equation}
Where c(t) is the time-dependent variation in count rate during the light curve, $c_{0}$ is the constant count rate during the quiescent state, $t_{0}$ and $t_{p}$ denote the flare start time and flare peak time, respectively, and $\tau_{r}$ and $\tau_{d}$ are the e-folding rise and decay times of the flare.\\
In the case of sets S2, S3, S4, and S5, the rotational modulation appears to be present as both pre-flare and post-flare states were observed at the different quiescent levels. Therefore, to account for the rotational modulation, we have fitted a sine wave with a period equal to the rotational period of AB Dor on the quiescent part of the light curve for sets S2, S4 and S5. However, in the case of set S3, we have fitted a sine wave with period half of the rotational period of AB Dor, which can occur due to the presence of two persistent active longitudes separated by 180$^\circ$ \citep[e.g.][]{2003A&A...405.1121B}. After removing the rotational modulation from the original light curve, we applied the mentioned flare model fitting. Figures \ref{fig:rot_mod} display the fitted sine wave, the original light curves, and the fitted flare model applied to the residual light curves. In the instance of set S1, we could not confirm the presence of rotational modulation, likely because the quiescent state was absent due to frequent flaring events occurring throughout the observation. Conversely, in the case of S6, significant rotational modulation was not detected.

We have used the data obtained from the PN detector to model the light curves for sets S1 and S2. However, RGS data was used for sets S3, S4, S5, and S6 because the PN detectors do not cover the entire observation for these sets.
The best-fit model parameters for all the 21 detected flares are shown in Table \ref{tab:lc_fitting}. The strength of these flares can be inferred from the quiescent to peak count rate ratio as mentioned in Table \ref{tab:lc_fitting} by the parameter F/Q. The observed values of F/Q in AB Dor vary from $\sim$1.2 to $\sim$34, with most flares having F/Q of 2 -- 4.

\begin{table*}
    \centering
    \caption{The best-fit model parameters for all the detected flares. Here F/Q represents the ratio of the flare peak count rate to the quiescent count rate. }
    \begin{tabular}{clll|clll}
         \hline
         Parameters($\rightarrow$)& $\tau_{r}$ & $\tau_{d}$ & F/Q & Parameters($\rightarrow$) & $\tau_{r}$ & $\tau_{d}$ & F/Q \\
         %\cline{2-10}
         Segments($\downarrow$)  & (ks) & (ks) & &
         Segments($\downarrow$)  & (ks) & (ks) & \\
         \hline %\lx $E_{flare}$ ($10^{28} erg s^{-1} $) ($10^{32} erg $) $L_{10}$ cm
         S1-F1  &  0.27$\pm{0.01}$ & 3.01$\pm{0.08}$ & 2.34$\pm{0.04}$ & ~~~F12 & 1.1$\pm{0.1}$   & 1.3 $\pm{0.2} $ &  2.00$\pm{0.09}$\\
         ~~~F2  &  0.43$\pm{0.06}$ & 3.7 $\pm{0.2} $ & 1.93$\pm{0.03}$ & ~~~F13 & 0.50$\pm{0.05}$ & 1.02$\pm{0.07}$ &  1.89$\pm{0.09}$\\
         ~~~F3  &  0.83$\pm{0.04}$ & 6.7 $\pm{0.2} $ & 1.59$\pm{0.03}$ & ~~~F14 &  1.2$\pm{0.2}$  & 2.9 $\pm{0.3} $ &  1.49$\pm{0.08}$\\
         S2-F4  &  0.35$\pm{0.05}$ & 0.9 $\pm{0.1} $ & 1.30$\pm{0.03}$ & S5-F15 & 0.52$\pm{0.02}$ & 2.19$\pm{0.05}$ &  34.4$\pm{0.4}$\\
         ~~~F5  &  0.8 $\pm{0.1} $ & 2.2 $\pm{0.2} $ & 1.16$\pm{0.03}$ & ~~~F16 & 0.66$\pm{0.06}$ & 0.92$\pm{0.04}$ &  3.5$\pm{0.1}$\\
         ~~~F6  &  0.36$\pm{0.02}$ & 0.84$\pm{0.05}$ & 1.60$\pm{0.03}$ & S6-F17 & 0.8$\pm{0.1}$   & 1.97$\pm{0.20}$ &  1.53$\pm{0.09}$ \\
         ~~~F7  &  2.19$\pm{0.08}$ & 2.9 $\pm{0.1} $ & 2.56$\pm{0.04}$ & ~~~F18 & 0.84$\pm{0.07}$ & 2.5 $\pm{0.2} $ &  1.79$\pm{0.09}$ \\
         ~~~F8  &  0.42$\pm{0.08}$ & 0.7 $\pm{0.2} $ & 2.84$\pm{0.05}$ & ~~~F19 & 0.51$\pm{0.04}$ & 1.9 $\pm{0.1} $ &   2.3$\pm{0.1}$\\ %F19 & $\sim$4.1 & 0.7$\pm{0.2}$ & 1.1$\pm{0.2}$ & 1.31$\pm{0.08}$ 
         ~~~F9  &  2.99$\pm{0.08}$ & 6.0 $\pm{1.0} $ & 3.50$\pm{0.05}$ & ~~~F20 & 0.96$\pm{0.02}$ & 4.39$\pm{0.06}$ &  11.0$\pm{0.2}$ \\
         S3-F10 &  0.36$\pm{0.05}$ & 1.09$\pm{0.09}$ & 2.00$\pm{0.05}$ & ~~~F21 & 4.9$\pm{0.9}$   & 2.4 $\pm{0.3} $ &   1.9$\pm{0.1}$ \\
         S4-F11 &  2.9 $\pm{0.7} $ & 1.28$\pm{0.09}$ & 1.89$\pm{0.09}$ &  \\
        % F12 & & 1.12$\pm{0.11}$ & 2.36$\pm{0.06}$ &&&1.84$\pm{0.1}$\\
        % F13 & & 0.96$\pm{0.02}$ & 4.39$\pm{0.06}$ & & & $4.4\pm0.5$\\
         \hline
    \end{tabular}
    %\caption{Caption}
    \label{tab:lc_fitting}
\end{table*}

\subsection{ EPIC Spectral analysis}
\label{sec:TRS}
In this section, we delve into an extensive discussion of the spectral analysis conducted on X-ray data from AB Dor, using observations from XMM-Newton. We employed time-resolved spectroscopy (TRS) to monitor changes in the X-ray spectral parameters throughout the observations. To begin, we isolated the flare components from the rest of the light curves. From the remaining light curve, we selected segments with the lowest mean count rates as pre- and/or post-flare segments, enabling us to identify the true quiescent state.

Our selection of 13 flares (F1-10, F13, F15, and F20 from RGS) out of the 21 total flares for time-resolved spectroscopy depended on the data available from the PN detector, which provided better statistics for low-exposure spectra. Furthermore, each flare was subdivided into multiple time bins, ensuring that each bin contained a similar and adequate number of counts for the subsequent spectral analysis.

\subsubsection{The Quiescent Spectroscopy}
\label{sec:TRS_qui}
\begin{figure}
    \centering
    \includegraphics[width=\columnwidth, trim={0.9cm 0.6cm 5.0cm 2.0cm},clip]{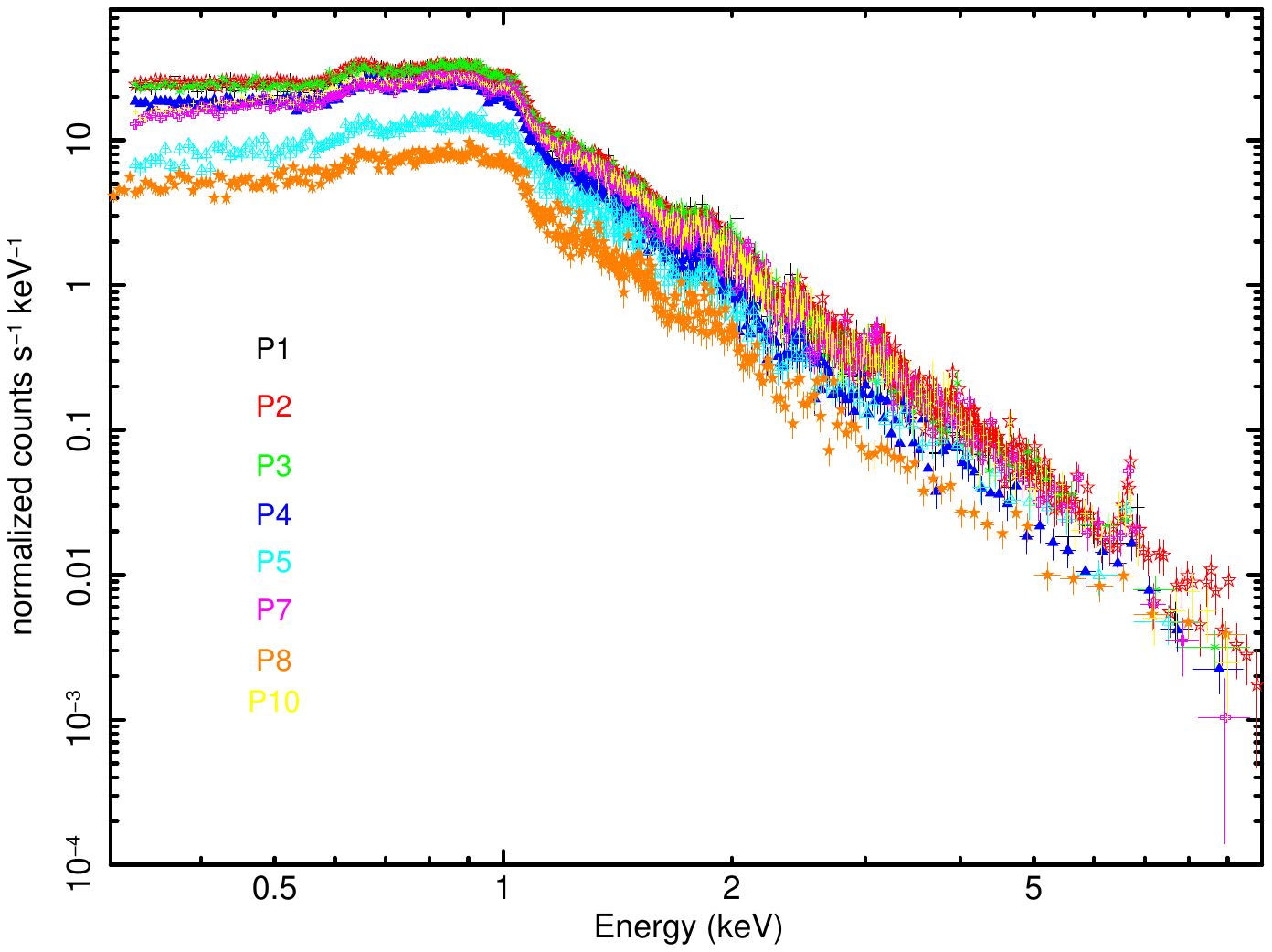}
    \caption{The X-ray spectra of pre- and post-flare states of AB Dor as obtained from EPIC-PN detector.}
    \label{fig:pn_all}
\end{figure}

We began by selecting the pre-/post-flare segments as proxies for the quiescent state in each observation and extracted the corresponding spectra. Figure \ref{fig:pn_all} showcases the quiescent state spectra of AB Dor from various observations. These X-ray spectra exhibited variations not only between different observations but also within a single observation. For instance, P3 and P4, both from the same observation, demonstrated inconsistency (refer also to Table \ref{tab:3apec_qui_all}).

To comprehend these variations, we conducted a comprehensive X-ray spectral analysis to determine parameters such as plasma temperature, emission measure, and abundance during these events. The quiescent state spectra were subjected to fitting with one (1-T), two (2-T), and three (3-T) temperature plasma models using {\sc apec} \citep[][]{2001ApJ...556L..91S}. We incorporated the X-ray absorption model {\sc phabs} to account for hydrogen column density (\nh), while solar photospheric abundances ($Z_{\odot}$) were adopted from \cite{1989GeCoA..53..197A}. During spectral fitting, all temperatures, emission measures, and global abundances (Z) remained free parameters.

For \nh, we fixed it at the maximum value of galactic \nh, which is $2\times10^{18}$ cm$^{-2}$ for AB Dor. This value of \nh was computed based on the maximum E(B-V) value of 0.0003 mag for AB Dor \cite[][]{2018A&A...616A.108B} using the relation from \cite{1975ApJ...198...95G}. The 3-T model exhibited a significantly better fit than the 1-T and 2-T models. The addition of another thermal component did not yield further improvement in the $\chi^{2}$, and the parameters of the fourth component were not well constrained.

Table \ref{tab:3apec_qui_all} summarizes the best-fit model parameters along with their reduced $\chi^{2}$ values for all the quiescent state spectra from P1 to P10. We calculated the average values of temperature (T${QA}$) and emission measures (EM${QA}$) for each segment using the following formulas:

\begin{equation}
T_{QA} = \frac{\sum_{i=1}^N T_{i} EM_{i}}{\sum_{i=1}^N EM_{i}} ; \quad EM_{QA} = \frac{1}{N}\sum_{i=1}^N EM_{i},
\end{equation}

Where $N$ (=3) represents the number of plasma components in the 3T models. The calculated average values of T${QA}$ and EM${QA}$ were 0.94 $\pm$ 0.06 keV and 4.6 $\pm$ 0.7 $\times$ $10^{52}$ \density, respectively. These average values exhibited consistency within a 2 $\sigma$ level (refer to Table \ref{tab:3apec_qui_all}). The X-ray luminosity (\lxq) was found to be variable, indicating varying coronal active regions from one observation to another.

\begin{table*}
    \centering
    \caption{Best fit spectral parameters from the pre/post-flare spectra using 3-T APEC model for all data sets.}
    \begin{tabular}{lccccccccccc}
         \hline
         Para ($\rightarrow$)& $kT_{1}$ & $kT_{2}$ & $kT_{3}$ & $T_{QA}$ & $EM_{1}$ & $EM_{2}$ & $EM_{3}$ & $EM_{QA}$ & Z  & \lxq  & $\chi_\nu^2 $ (dof) \\
         %\cline{2-10}
         %Segment($\downarrow$) & (keV) & (keV) & (keV) &(10$^{52}$ \density)&(10$^{52}$ \density)&(10$^{52}$ \density)& (Z$_\odot$) & (10$^{20}$ cm$^{-2}$) & ($10^{30}$ \lum)\\
         Seg ($\downarrow$) &  &  &  &  &  &  & & & (Z$_\odot$) &    \\
         \hline
         P1 & 0.297$_{-0.006}^{+0.006}$ & 0.98$_{-0.01}^{+0.01}$ & 2.2$_{-0.2}^{+0.2}$ &  0.90$_{-0.04}^{+0.05}$ & 5.4$_{-0.2}^{+0.2}$ & 6.6$_{-0.5}^{+0.5}$ & 2.1$_{-0.3}^{+0.4}$ & 4.7$_{-0.2}^{+0.2}$ & 0.18$_{-0.01}^{+0.01}$ & 1.165$_{-0.005}^{+0.005}$ & 1.27 (403)\\
         \\
         P2 & 0.288$_{-0.002}^{+0.003}$ & 0.974$_{-0.005}^{+0.006}$ & 1.99$_{-0.04}^{+0.04}$ & 1.0$_{-0.01}^{+0.01}$ & 4.8$_{-0.1}^{+0.1}$ & 6.3$_{-0.2}^{+0.2}$ & 3.6$_{-0.1}^{+0.1}$ & 4.90$_{-0.08}^{+0.08}$ & 0.194$_{-0.005}^{+0.005}$ & 1.277$_{-0.002}^{+0.002}$ & 1.79 (742)\\
         \\
         P3 & 0.287$_{-0.007}^{+0.008}$ & 0.97$_{-0.01}^{+0.01}$ & 2.1$_{-0.1}^{+0.3}$ & 0.95$_{-0.04}^{+0.06}$ & 4.6$_{-0.2}^{+0.2}$ & 6.9$_{-0.4}^{+0.6}$ & 2.5$_{-0.4}^{+0.3}$ & 4.7$_{-0.2}^{+0.2}$ & 0.19$_{-0.01}^{+0.01}$ & 1.214$_{-0.005}^{+0.005}$ & 1.03 (434)\\
         \\
         P4 & 0.282$_{-0.006}^{+0.007}$ & 0.94$_{-0.02}^{+0.02}$ & 1.8$_{-0.1}^{+0.2}$ & 0.85$_{-0.04}^{+0.05}$ &3.7$_{-0.2}^{+0.2}$ & 4.0$_{-0.3}^{+0.4}$ & 1.8$_{-0.3}^{+0.3}$ & 3.2$_{-0.2}^{+0.2}$ & 0.20$_{-0.01}^{+0.02}$ & 0.863$_{-0.004}^{+0.004}$ & 1.54 (361)\\
         \\
         P5 & 0.298$_{-0.008}^{+0.01}$ & 0.96$_{-0.02}^{+0.02}$ & 2.0$_{-0.1}^{+0.1}$ & 1.03$_{-0.04}^{+0.04}$ &4.3$_{-0.3}^{+0.3}$ & 5.7$_{-0.5}^{+0.6}$ & 3.7$_{-0.4}^{+0.4}$ &  4.6$_{-0.2}^{+0.3}$ & 0.21$_{-0.02}^{+0.02}$ &  1.251$_{-0.007}^{+0.007}$ & 1.07 (356)\\
         \\
         P7 & 0.278$_{-0.005}^{+0.005}$ & 0.86$_{-0.02}^{+0.02}$ & 1.5$_{-0.1}^{+0.1}$ & 0.97$_{-0.05}^{+0.05}$ & 3.5$_{-0.2}^{+0.2}$ & 3.6$_{-0.3}^{+0.3}$ & 5.3$_{-0.3}^{+0.3}$ & 4.1$_{-0.2}^{+0.2}$ & 0.23$_{-0.01}^{+0.02}$ & 1.07$_{-0.01}^{+0.01}$ & 1.68 (523) \\
         \\
         P8 & 0.282$_{-0.007}^{+0.006}$ & 0.97$_{-0.01}^{+0.01}$ & 1.9$_{-0.1}^{+0.2}$ & 0.89$_{-0.03}^{+0.05}$ & 5.8$_{-0.3}^{+0.3}$ & 7.4$_{-0.6}^{+0.6}$ & 2.9$_{-0.4}^{+0.4}$ & 5.4$_{-0.3}^{+0.3}$ & 0.21$_{-0.01}^{+0.01}$ & 1.411$_{-0.006}^{+0.006}$ & 1.29 (412)\\
         \\
         P10 & 0.270$_{-0.004}^{+0.004}$ & 0.952$_{-0.008}^{+0.008}$ & 1.9$_{-0.1}^{+0.1}$ & 0.91$_{-0.03}^{+0.03}$ & 5.9$_{-0.2}^{+0.2}$ & 6.9$_{-0.3}^{+0.3}$ & 3.6$_{-0.2}^{+0.2}$ & 5.5$_{-0.1}^{+0.1}$ & 0.18$_{-0.01}^{+0.01}$ & 1.346$_{-0.004}^{+0.004}$ & 1.28 (546)\\
         \\
         Avg & 0.29$\pm$0.01 & 0.95$\pm$0.04 & 1.9$\pm$0.2 & 0.94$\pm$0.06 & 4.8$\pm$0.9 & 6$\pm$1 & 3$\pm$1 & 4.6$\pm$0.7 & 0.20$\pm$0.02 &  1.200$\pm$0.2\\
         \hline
    \end{tabular}
    ~~\\
     Note: The first row and first column represent the parameters (para) and light curve segments (seg). 
     All the temperatures and emission measures are in units of keV and 10$^{52}$ \density, respectively, whereas X-ray luminosity (\lxq) is in units of $10^{30}$ \lum. The error in the average values of the parameters in the last row is the standard deviation. The dof is the number of degrees of freedom, and $\chi_\nu^2$ is the reduced $\chi^2$.
    \label{tab:3apec_qui_all}
\end{table*}

\subsubsection{Spectral evolution during the flares}
\label{sec:TRS_flare}
%from the EPIC-PN instrument
To trace the evolution of spectral parameters during flares, we divided the full flare light curve into several time segments consisting of the rising, peak, and decay phases and generated spectra for each segment. The rising, peak and decay phases are denoted by Ri, P, and Di, respectively. Here i = 1,2..... The length of each segment was chosen in such a way that each segment contained an equal number of counts. The TRS was conducted on all available flares using the PN detector due to its high signal-to-noise ratio. Additionally, TRS was performed for flare F20 due to high count rates in the RGS spectra.

The X-ray spectra of flaring segments were also fitted with 3-T {\sc apec} model. All the parameters of 3-T {\sc apec} model were kept free during the spectral fitting. The first two temperatures were found to be constant within a 1$\sigma$ level and were similar to the quiescent temperatures. However, the third temperature was found to be varying during the flare. Therefore, for further spectral fitting, we have fixed the first two temperatures, and corresponding normalizations at the quiescent level and \nh at $2\times10^{18}$ cm$^{2}$. Other parameters like Z,  the temperature of the third component, and the corresponding normalization were kept free. 

\begin{figure*}
\centering
\vspace{-0.1cm}
\subfigure[set S1 ]{\includegraphics[width=0.9\columnwidth, trim={0.7cm 1.0cm 2.0cm 3.5cm}, clip]{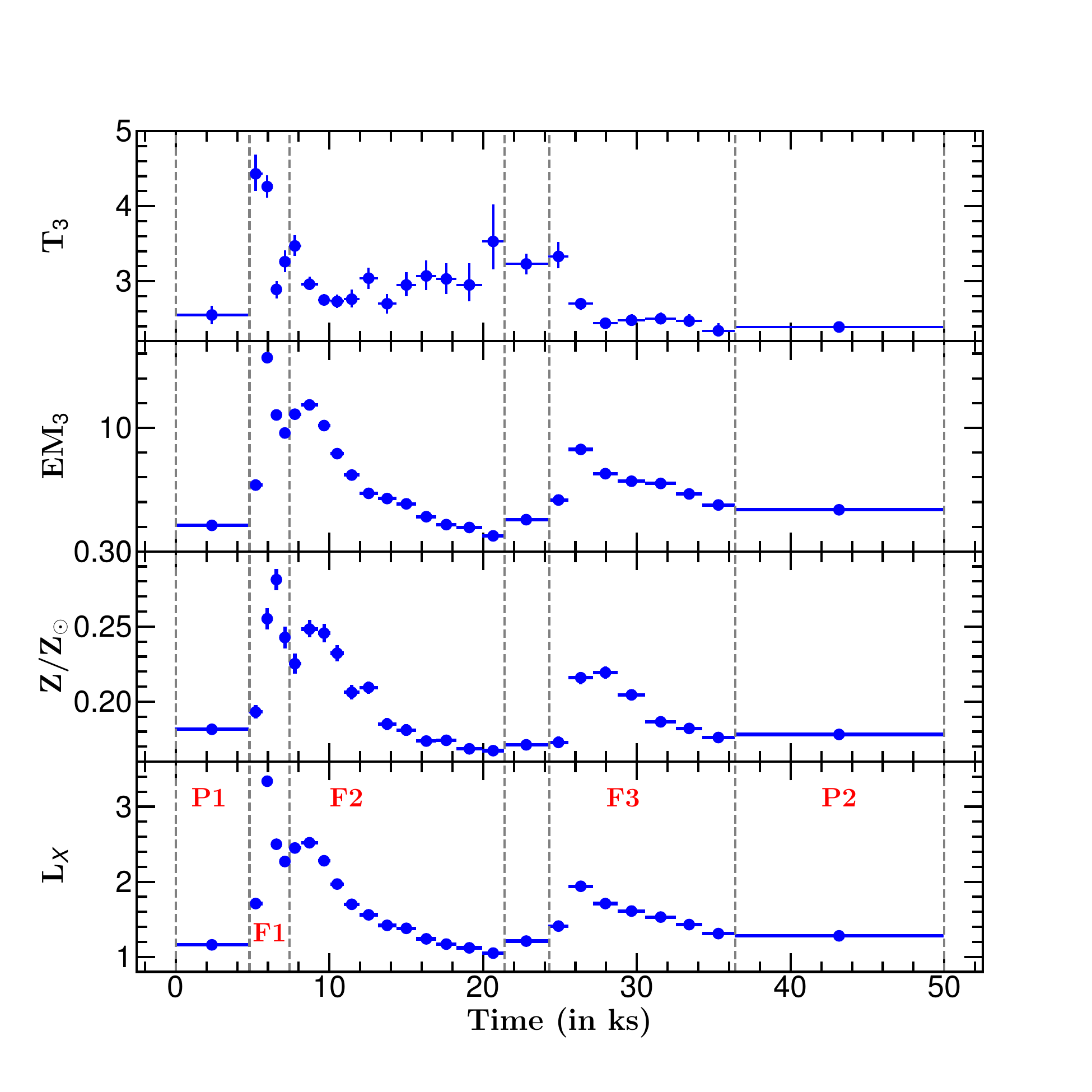}}
\subfigure[set S2 ]{\includegraphics[width=0.9\columnwidth,trim={0.7cm 1.0cm 2.0cm 3.5cm},clip]{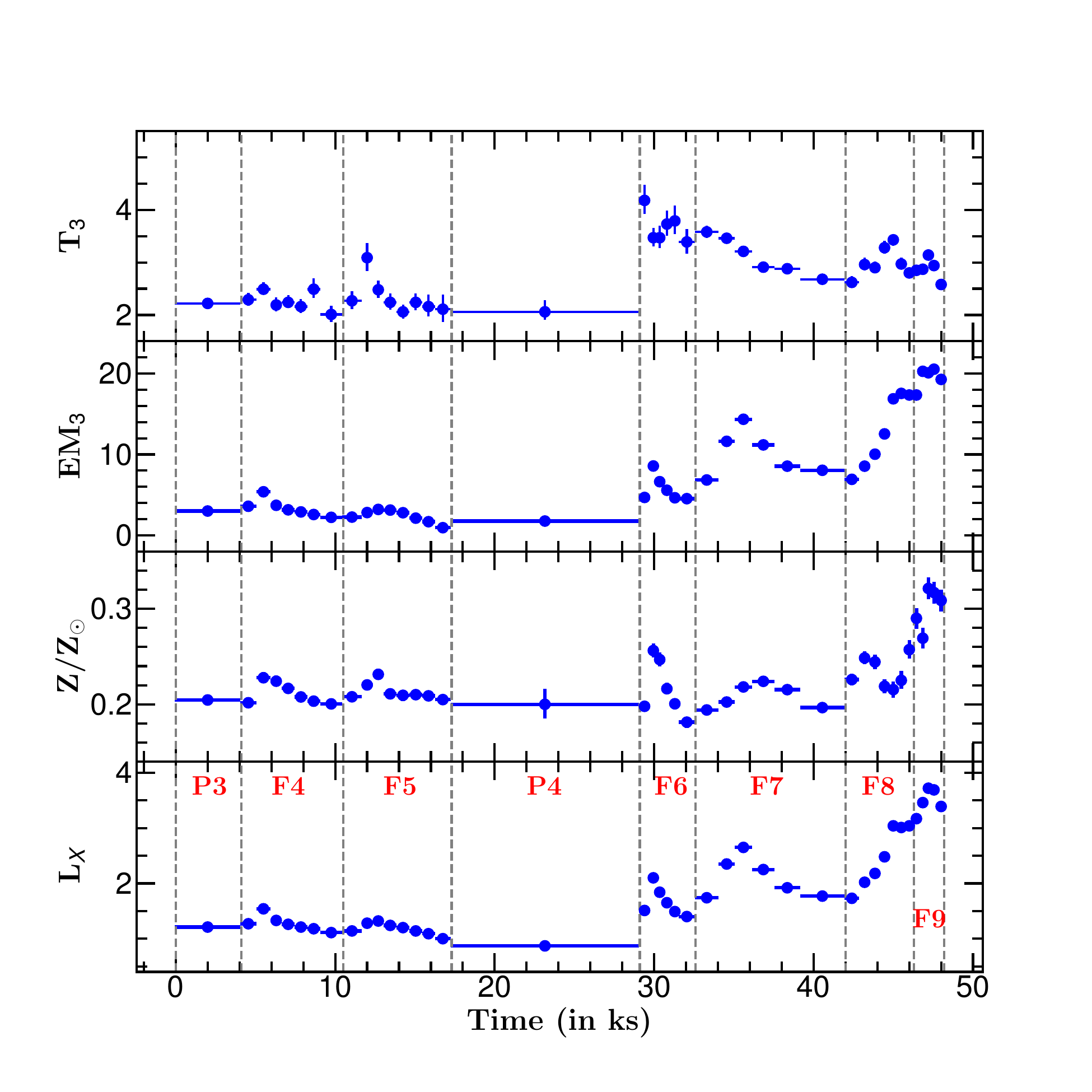}}
\vspace{-0.2cm}
\subfigure[set S3 ]{\includegraphics[width=0.9\columnwidth, trim={0.7cm 1.0cm 2.0cm 3.5cm},clip]{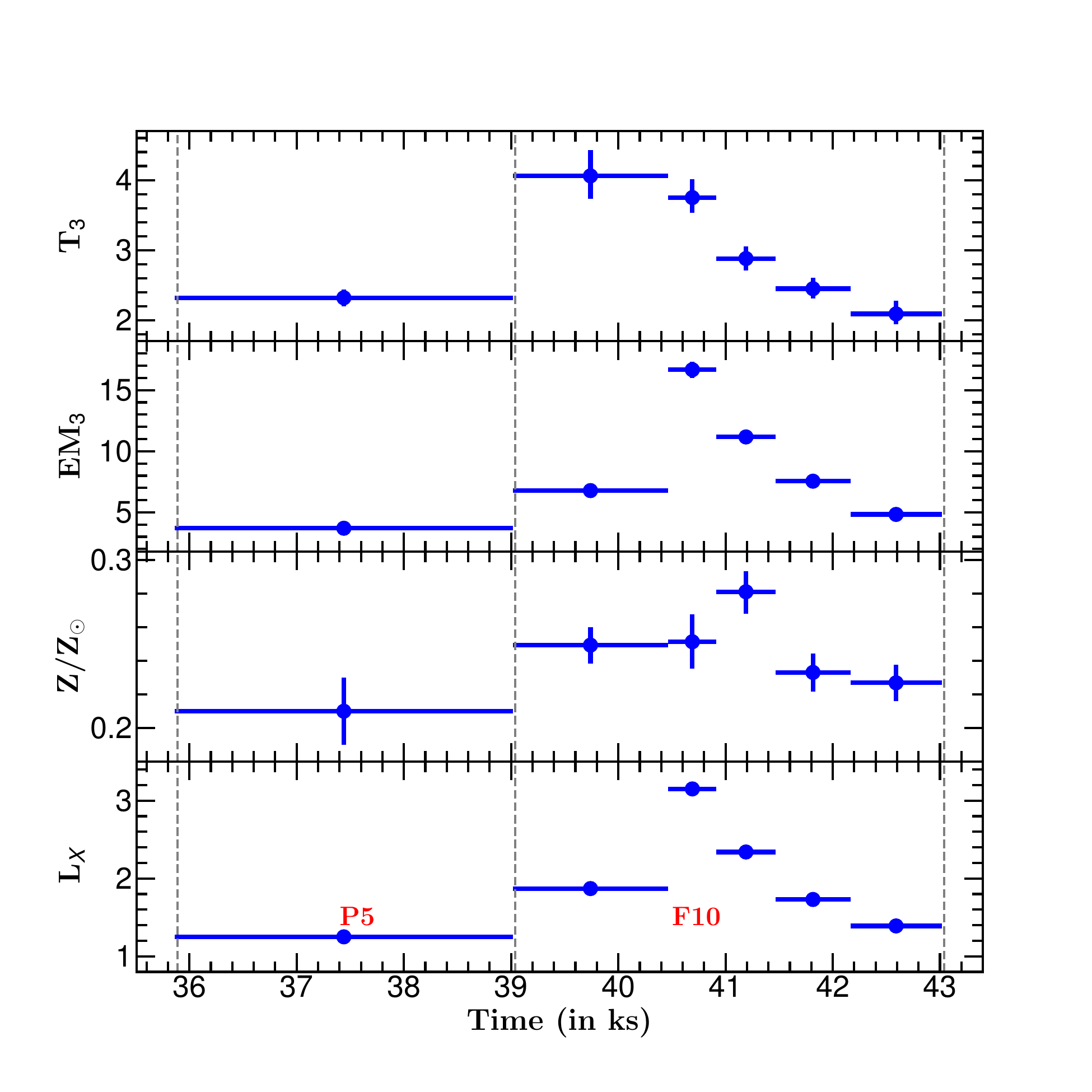}}
\subfigure[set S4 ]{\includegraphics[width=0.9\columnwidth,trim={0.7cm 1.0cm 2.0cm 3.5cm},clip]{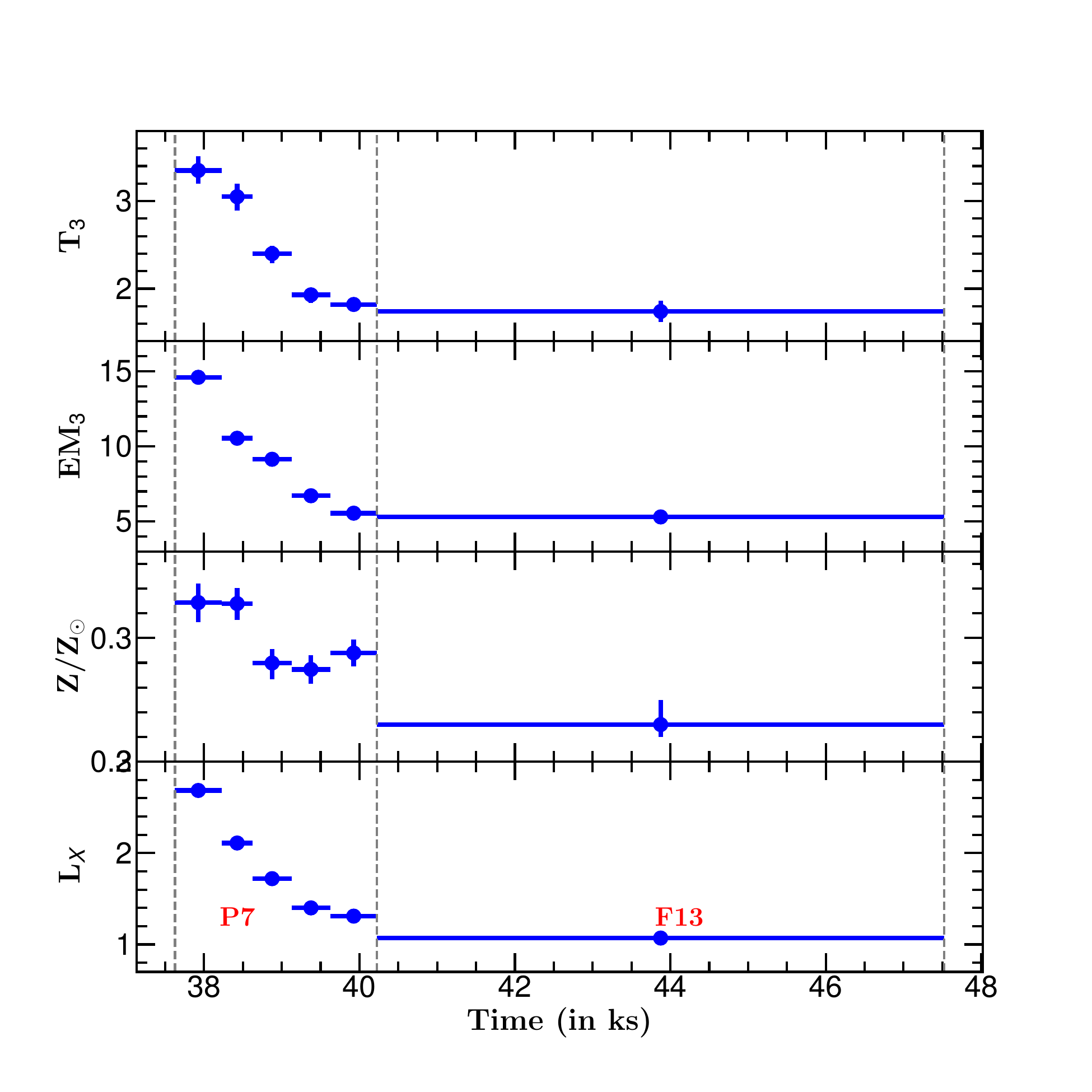}}
\vspace{-0.2cm}
\subfigure[set S5 ]{\includegraphics[width=0.9\columnwidth,trim={0.7cm 1.0cm 2.0cm 3.5cm},clip]{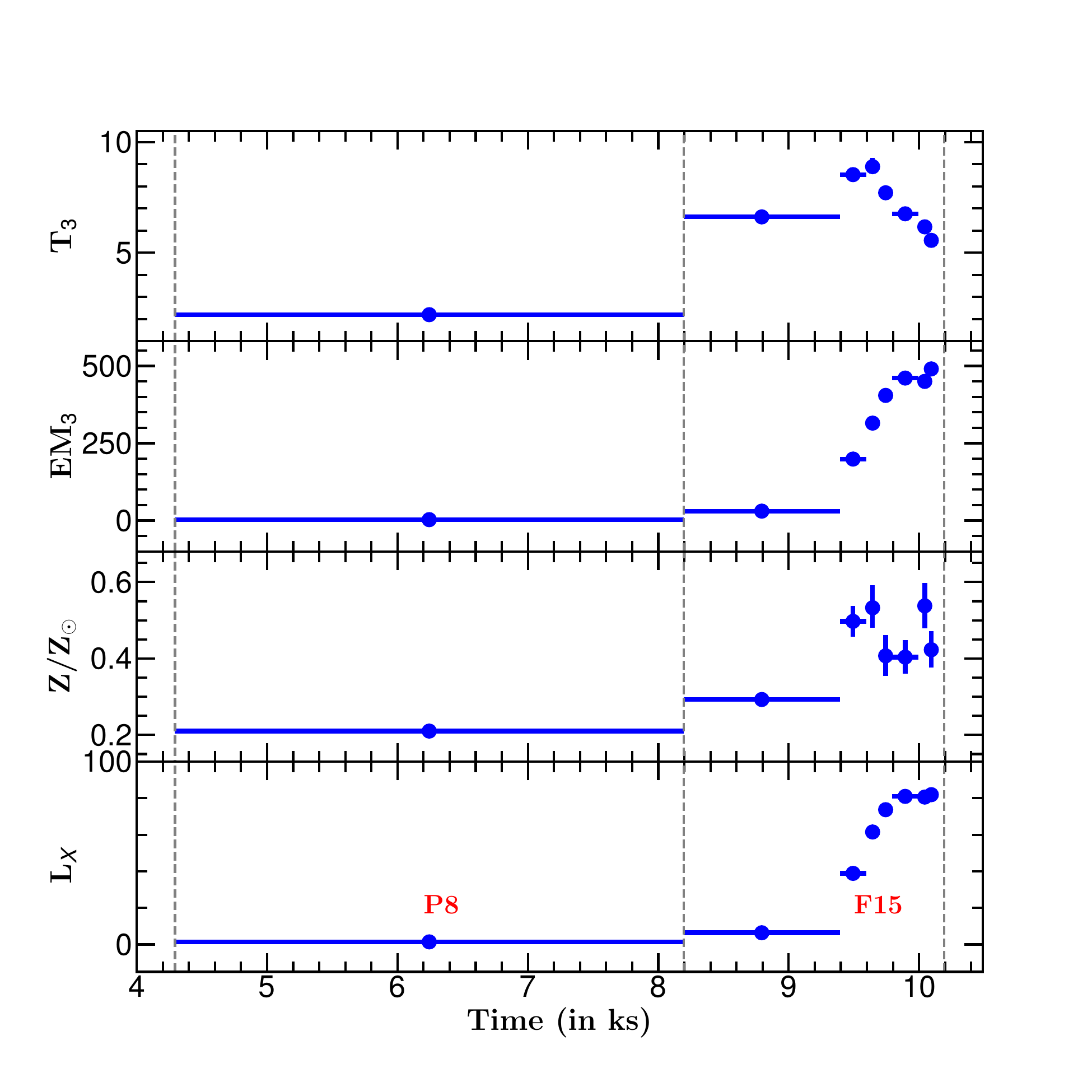}}
\subfigure[set S6 ]{\includegraphics[width=0.9\columnwidth,trim={0.7cm 1.0cm 2.0cm 3.5cm},clip]{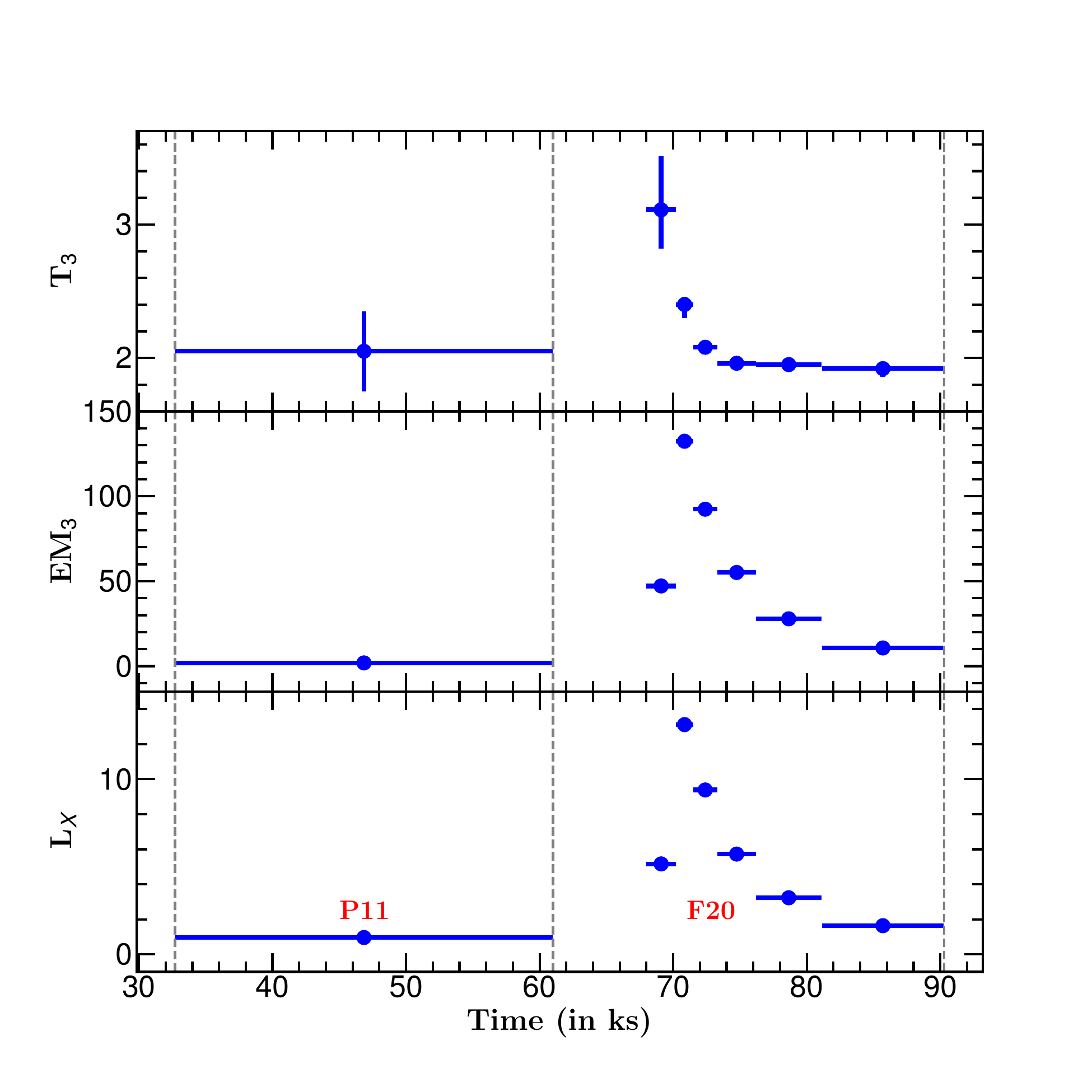}}
\vspace{-0.25cm}
\caption{Temporal evolution of the spectral parameters of AB Dor during flares and quiescent states, where top to bottom plots show the variation of temperature ($T_{3}$) in units of $10^{7}$ K, emission measure ($EM_{3}$) in units of $10^{52}$ \density, relative abundance (Z/Z$_\odot$), and X-ray luminosity in units of $10^{30}$ \lum. }
\label{fig:tvspara}
\end{figure*}

We have also fitted the spectra with the 3+1-T {\sc apec} model, fixing all the parameters of the first three components to the nearest quiescent state value and varied parameters of the fourth component. This model either does not constrain the model parameter or does not improve the statistics of spectral fit. Therefore, we opted for the previous model for further analysis. All the best fit spectral parameters with 68\% confidence range and reduced $\chi^{2}$ are given in Table \ref{tab:all_flares_2+1apec} and temporal variation of all these parameters is shown in Figure \ref{fig:tvspara}.

The top panels of Figure \ref{fig:tvspara} depict the temporal evolution of the third temperature T3, which shows that the flare temperature is maximum during the rising phase of the flare. The T3 reached a value of 7.7 keV during the flare F15, the highest of all the flares analysed here. The maximum flare temperature T3 ranges from 2.7 to 3.8 keV for the rest of the flares. Also, we have found in the case of multiple flares (flares F1 and F2 from set S1 and flares F6 to F9 from set S2), the temperature during the decay phase of the flare is increasing rather than decreasing. 
This discrepancy can be attributed to the overlap between the decay phase of the first flare and the rising phase of the second flare. Within this overlapping region, the heating generated during the ascending phase of the second flare influences the cooling or decay phase of the preceding flare. Consequently, this overlapping effect leads to an overall increase in temperature.
Further, the variation in emission measure EM$_3$, abundance Z, and X-ray luminosity \lxf are found to be in agreement with the standard flare model and are shown in the second, third, and fourth plots of Figures \ref{fig:tvspara} (a) to (f). The values of peak EM$_3$, Z, and \lxf were also found to be maximum for the flare F15 as 4.91 $\times$ $10^{54}$ \density, 0.54 Z$_\odot$, and >42.4 $\times$ 10$^{31}$ \lum, respectively, whereas for the remaining flares the peak EM$_3$, Z and L$_{XF}$ were found to be in the range of 3.2 $\times$ $10^{52}$ -- 1.3 $\times$ $10^{54}$ \density, 0.22 -- 0.33 Z$_\odot$, and 0.9 -- 3.6 $\times$ 10$^{31}$ \lum, respectively.

\begin{figure*}
    \centering
    \includegraphics[width=\linewidth,trim={0.0cm 1.0cm 5.0cm 2.0cm}, clip]{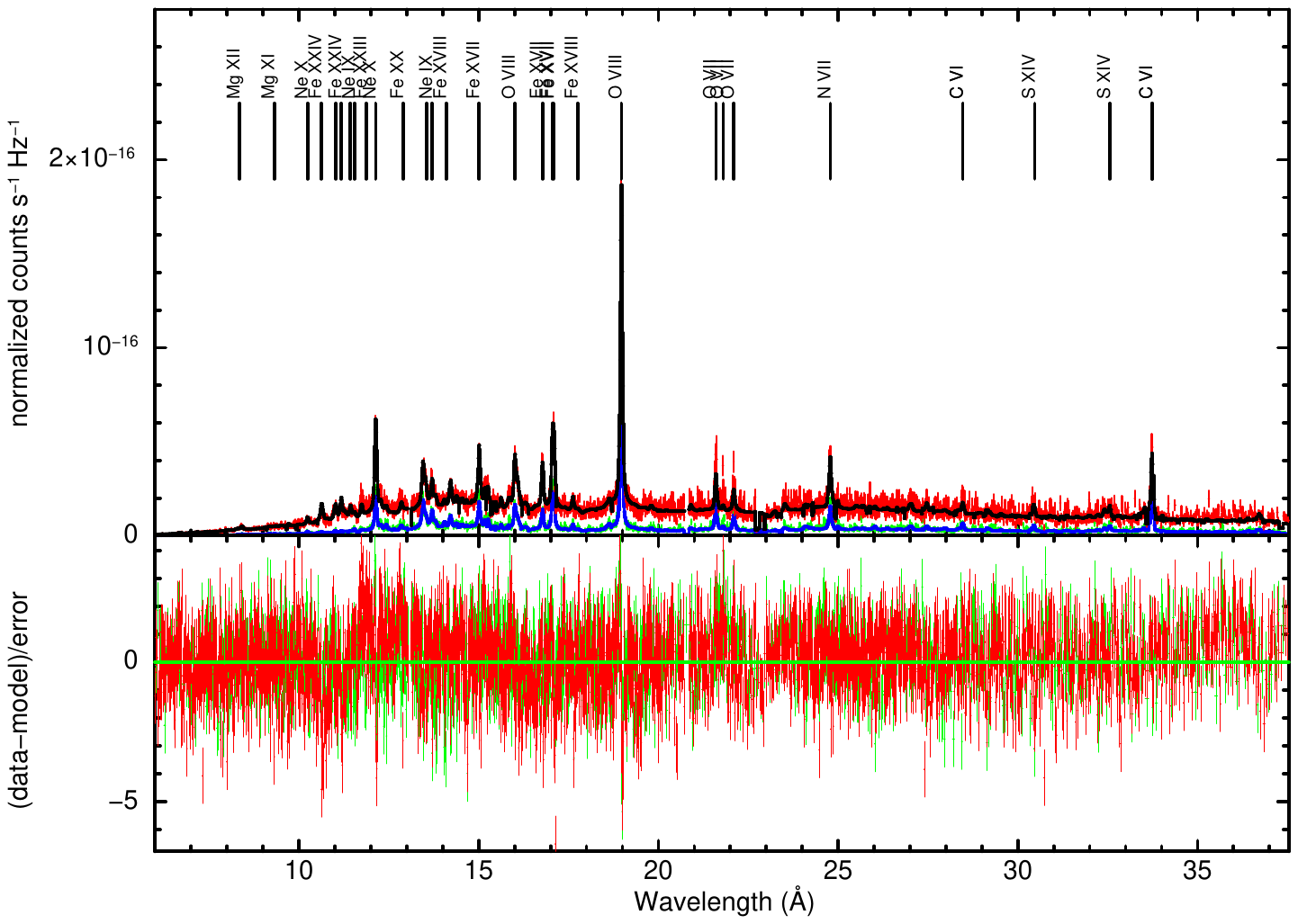}
    \caption{RGS spectra during quiescent (P11) and flaring state (F20) from the set S6 are shown in green and red color, repectively. The best fit 3-temperature {\sc vapec} model for both are shown in blue and black solid lines, respectively. }
    \label{fig:RGS_fullspectra}
\end{figure*}

\subsection{RGS Spectral Analysis}
\label{sec:RGS_spectra}
The RGS spectra were generated for both quiescent and flaring states and fitted with the 3-T {\sc vapec} model along with the model {\sc phabs} to account for the \nh. For the quiescent states spectral fitting, the temperatures, and emission measures were kept free, whereas the \nh was fixed at $2\times10^{18}$ cm$^{2}$. However, the abundances of Al, and Ni were fixed to the solar photospheric values, while the remaining abundances were kept free and tied among each temperature component. 

For the spectral fitting of the flaring states, we applied a similar approach as applied for the PN-spectral fitting. However, we used 3-T {\sc vapec} model instead of the 3-T {\sc apec} model. We used the first two temperatures (kT$_1$ and kT$_2$) and emission measures (EM$_1$ and EM$_2$) as the proxy of the quiescent state. In order to achieve a sufficient signal-to-noise ratio, the flare in RGS requires stronger re-binning for spectral analysis than for the PN. Therefore, rather than dividing the flare into different segments as done for the PN spectral analysis, we used the spectra of the entire flaring event to determine the spectral parameters. 
   
The best fit model parameters within a 68\% confidence range are given in Table \ref{tab:3vapec_qui_all} for both quiescent and flaring states of AB Dor. The best-fit preflare P11 and flare F20 spectra with the residual of the best-fit 3-T {\sc vapec} model are also illustrated in Figure \ref{fig:RGS_fullspectra}.

\begin{figure*}
\centering
%\vspace{-1.0cm}
\subfigure[Set 2]{\includegraphics[width=\columnwidth]{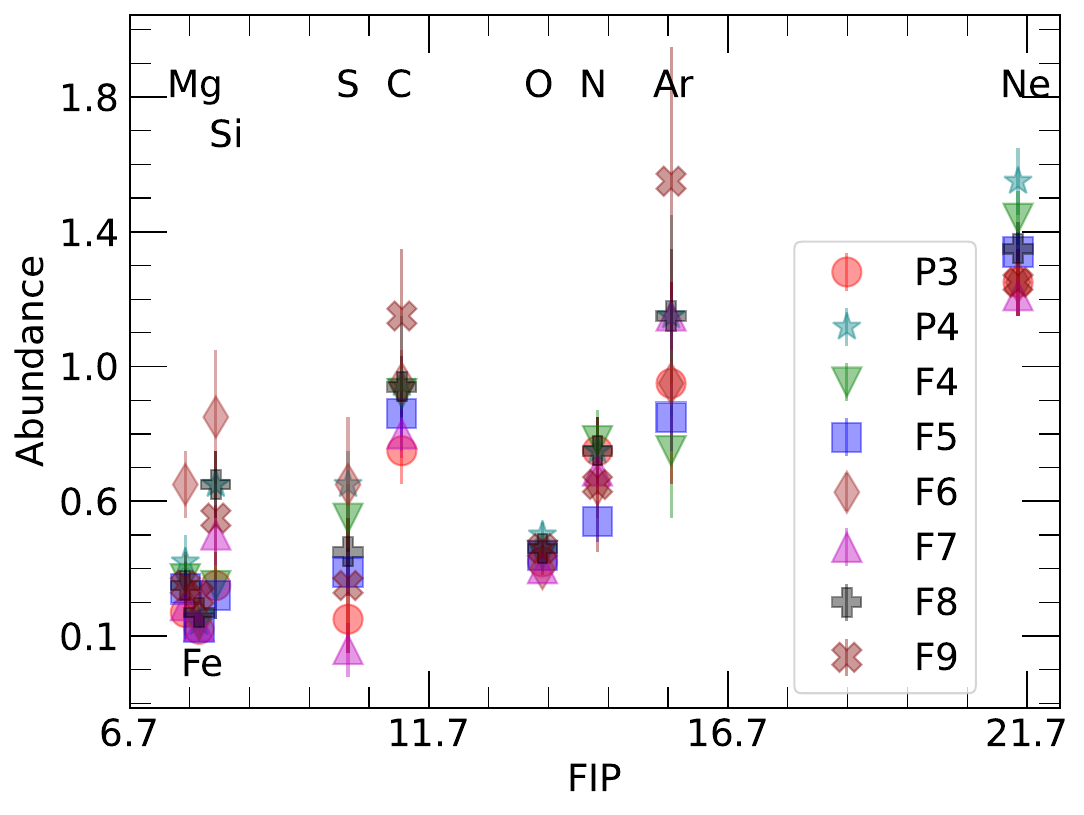}}
\subfigure[Set 3]{\includegraphics[width=\columnwidth]{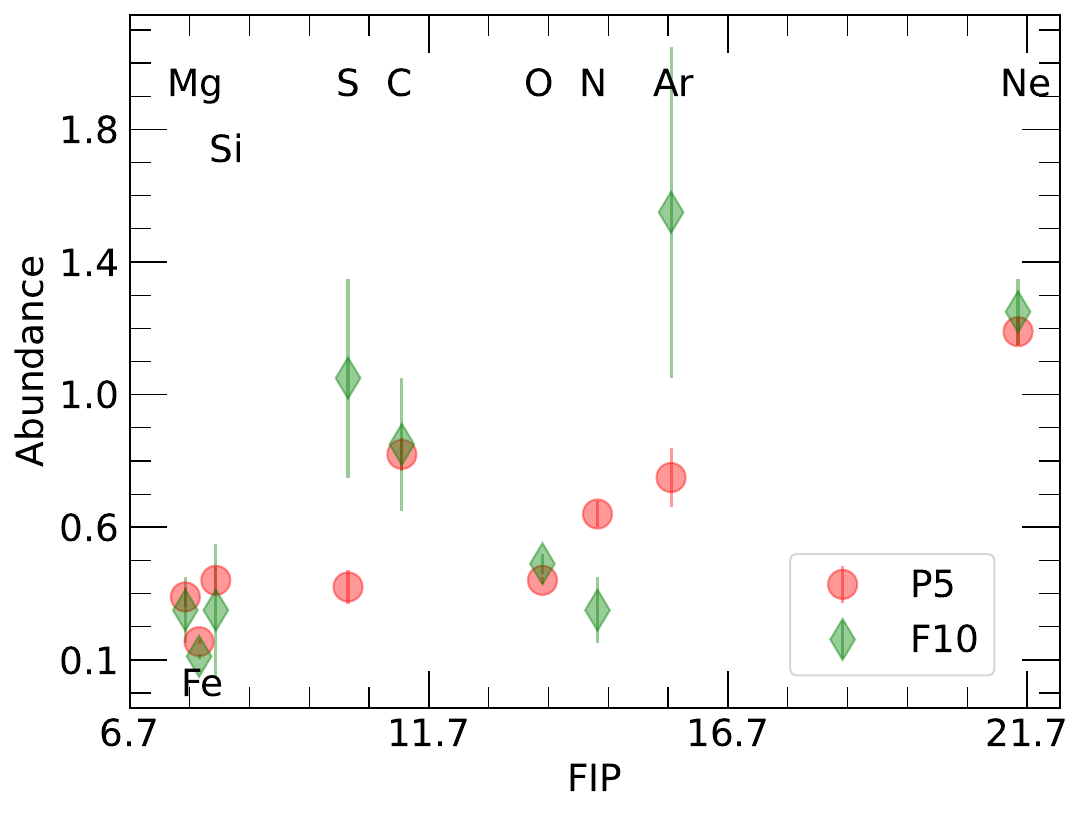}}
\subfigure[Set 4]{\includegraphics[width=\columnwidth]{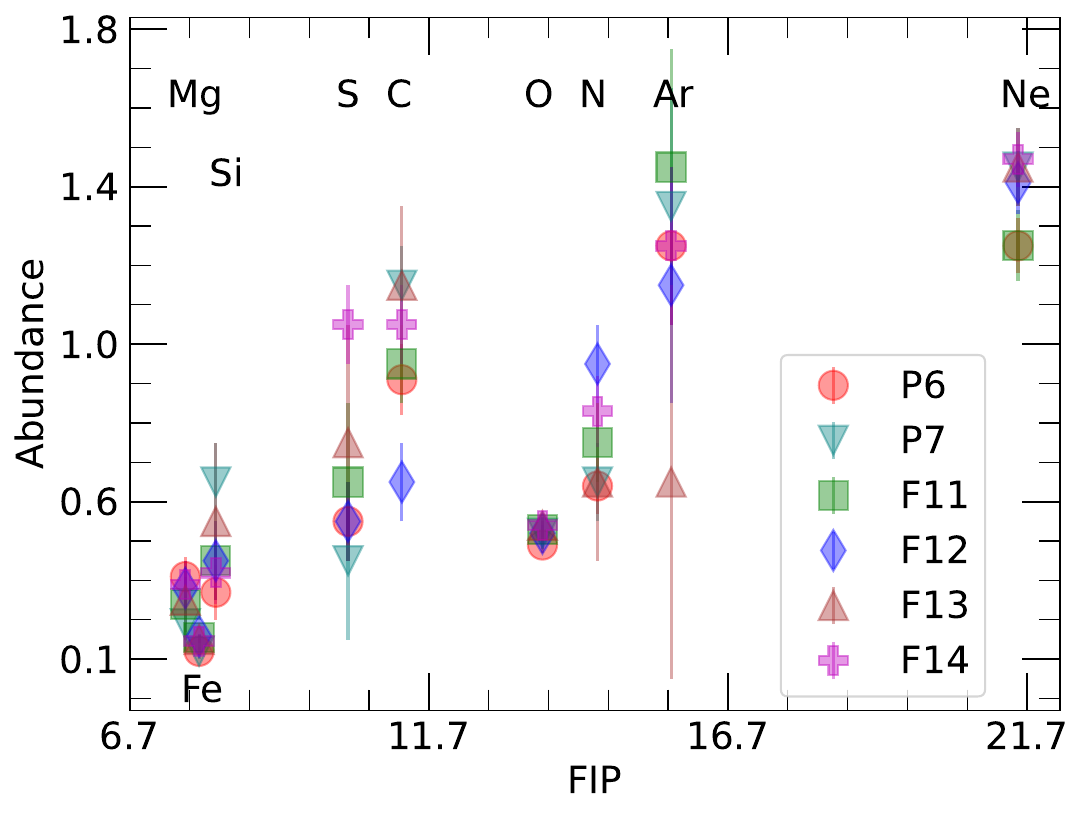}}
\subfigure[Set 5]{\includegraphics[width=\columnwidth]{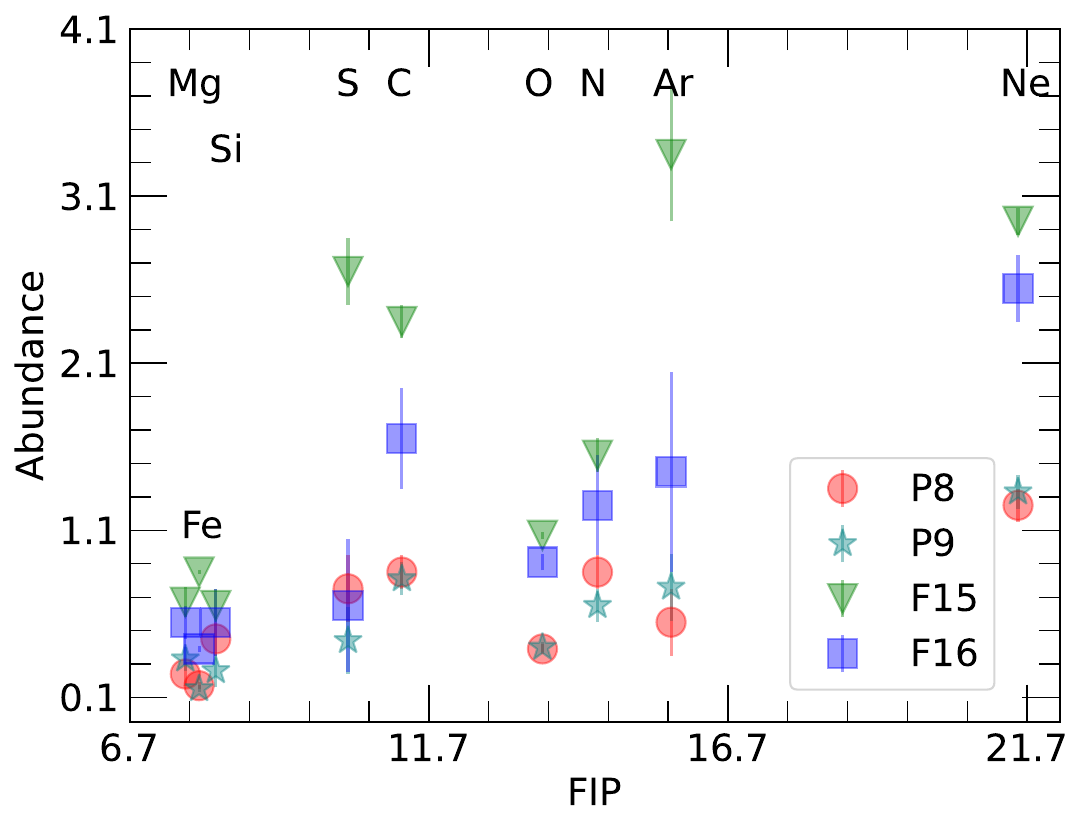}}
\subfigure[Set 6]{\includegraphics[width=\columnwidth]{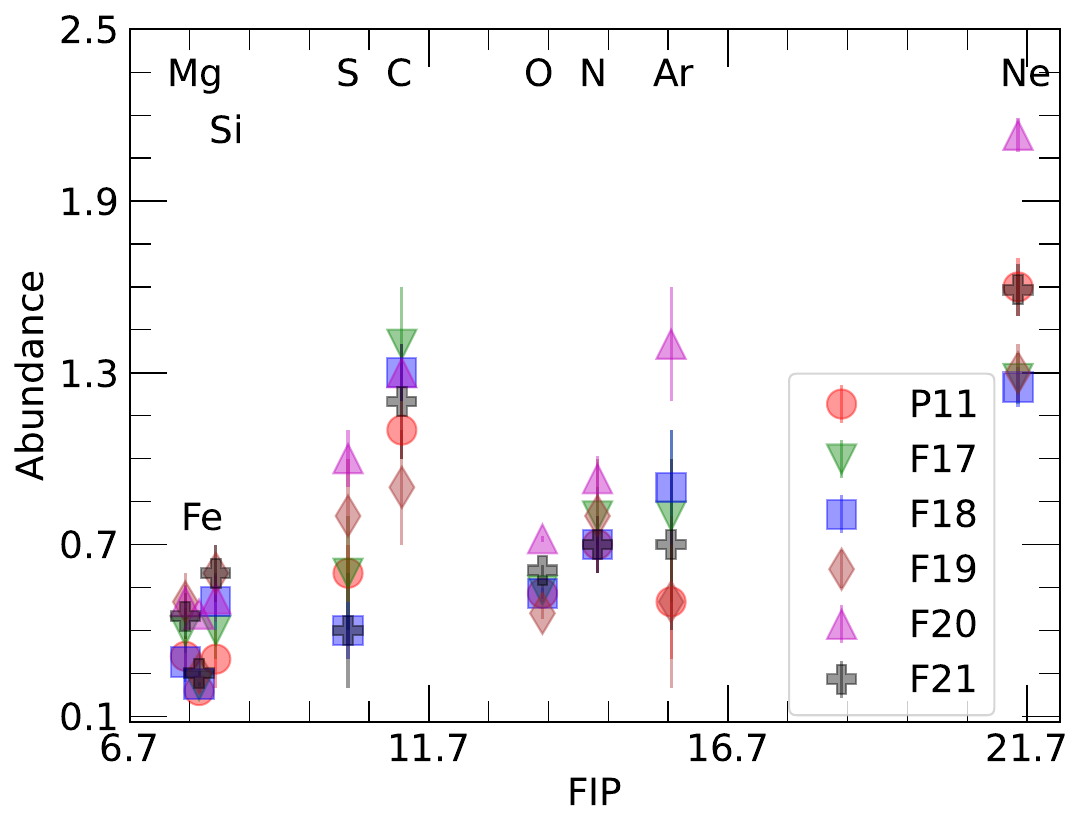}}
\caption{Elemental abundances plotted as a function of FIP for both quiescent and flaring states during different observations of AB Dor.} \label{fig:IFIP_plots_all}
\end{figure*}

\subsubsection{FIP and Inverse FIP effect}
\label{sec:IFIP}
%defination
The elemental abundances in the corona are not the same as those in the underlying photosphere because of the fractionation process related to the first ionization potential (FIP) of the elements. In slow rotators (e.g. Sun), generally, the so-called FIP effect is observed in which elements with FIP $<$ 10 eV are enhanced relative to the elements with FIP $>$ 10 eV \citep[][]{1992PhyS...46..202F,1995ApJ...443..416L,2000PhyS...61..222F}.
However, it has been shown for a few fast rotating stars that the Inverse-FIP (I-FIP) effect occurs due to their higher magnetic activity \citep[see][]{2015LRSP...12....2L,2021ApJ...909...17L}.
It has been observed that some stars do not show any FIP bias, which has been linked to their magnetically inactive chromospheres \citep[][]{1994AAS...184.0522D}.

Figure \ref{fig:IFIP_plots_all} shows the measured abundance as a function of FIP for the quiescent and flaring phases of different observations of AB Dor. The coronal abundances of elements Mg, Fe and Si with FIP <10 eV, are found to be under-abundant relative to solar photospheric values for both the quiescent and flaring states. The average abundance of Mg, Fe and Si during quiescent states are found to be $\sim$0.31, $\sim$0.19, and $\sim$0.42, respectively. Also, all other high FIP elements except Ne (with an average abundance of $\sim$ 1.34) are found to be underabundant during quiescent states. 
Furthermore, our observations revealed the presence of the Inverse First Ionization Potential (I-FIP) effect during both quiescent and flaring states. 
In the majority of the flaring periods, the abundances of individual elements were observed to be higher compared to those in the quiescent state.

\subsection{Loop modelling}
\label{sec:LLC}
Although stellar flares can't be resolved spatially, their analogies with solar flares and loop models help us to infer the morphology and physical size of flaring loops and stellar corona. 
Observations of stellar flares do not always cover the entire duration of flares. In some cases, the rising phase is not covered, and in other cases, the decay phase is not observed. Also, in some flares, there is an overlap between the decaying phase of one flare and the rising phase of the other. In our case, only the decay phase of flare F13 and the rising phase of flare F15 are observed with the PN detector. So, the separate calculations for loop length from flare rise \citep{2007A&A...471..271R} and decay phases \citep{1997A&A...325..782R} have been made using the Hydrodynamic loop model. This model assumes a single dominant coronal loop which includes plasma cooling as well as the heating effect during flare decay. We have determined the semi-loop length (L), using both of the approaches outlined below.

\subsubsection{The Rise phase}
\label{sec:LLC_rise}
\cite{2007A&A...471..271R} introduced a detailed model for loop length calculation using the rise and peak phases of the stellar X-ray flare.
As a consequence of the heating event, the loop's temperature experiences a rise, reaching its maximum at temperature $T_{0}$. Concurrently, due to chromospheric evaporation, the density also rises and peaks at temperature $T_{M}$ within the flaring loop. During the peak phase, heating stops and conduction cooling starts to dominate, which causes the temperature to drop at the peak phase. At this stage, the continuous increase in density indicates the ongoing evaporation process.

Using hydrodynamic simulations of semi-circular flaring loops with constant cross-section, \cite{2007A&A...471..271R} derived an empirical formula for semi-loop length as

\begin{equation}
\label{eq_rise}
{\rm L_{r} = 950 \frac{T_0^{5/2}}{T_M^2}t_{M} ~~cm~~ }
\end{equation}

Where L$_r$ is the semi-loop length in units of cm, T$_{0}$ and T$_{M}$ are the maximum temperature and temperature at maximum density, respectively, in units of K, and $t_{M}$ is the time at which density maximum occurs in units of sec. The derived semi-loop lengths using this model are given in Table \ref{tab:final_parameters} and were found to be in the range of 0.8 -- 4.5 $\times$ $10^{10}$ cm. The largest loop length was found for flare F15, whereas the smallest loop length was found for flares F1, F6, and F9.

\subsubsection{The decay phase}
\label{sec:LLC_decay}
\begin{figure}
    \includegraphics[width=\columnwidth]{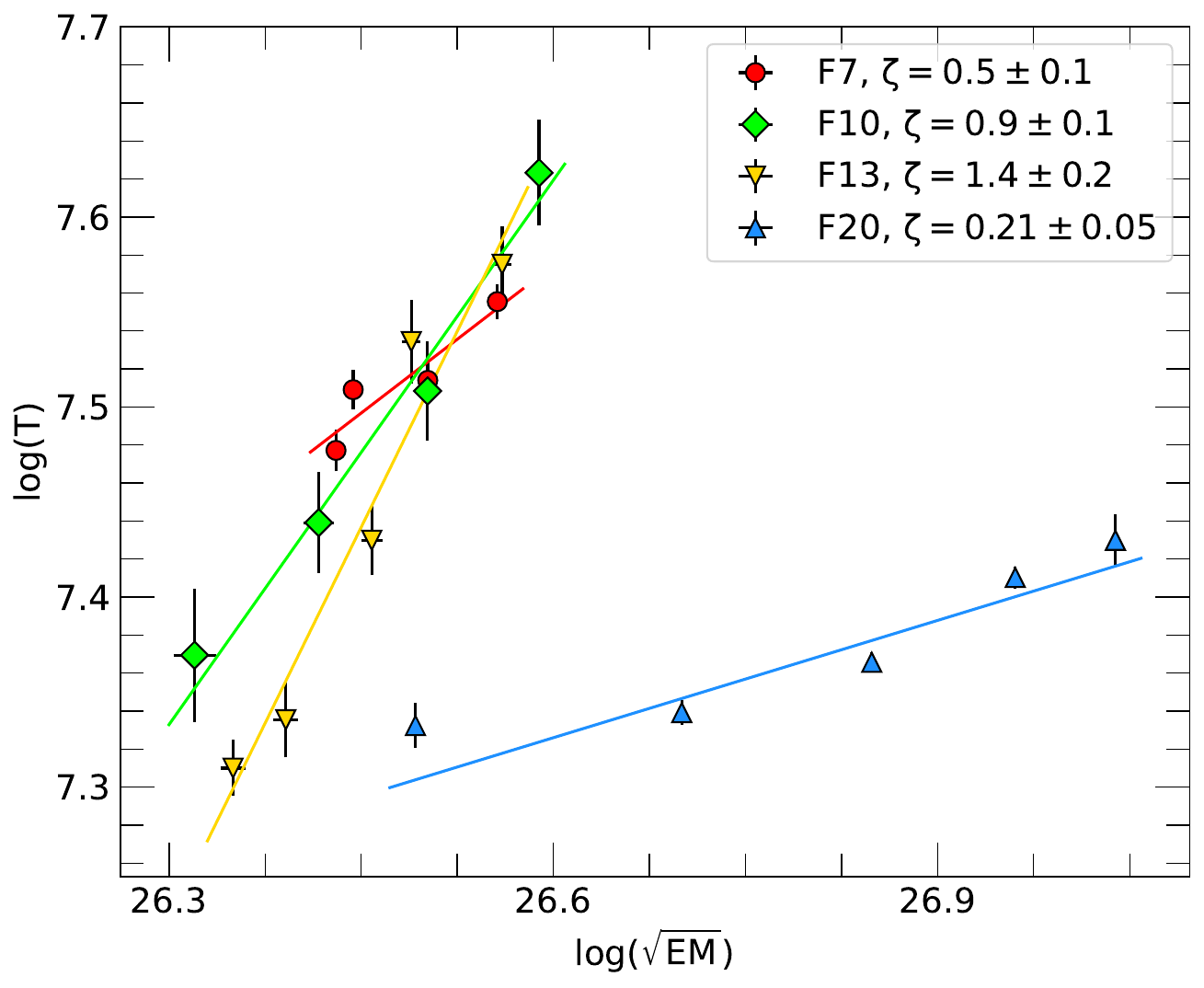}
\caption{Flare evolution in the log$\sqrt{EM}$ versus log(T) curve during the decay phase for flares F7, F10, F13, and F20 along with the best fit straight line. The slope ($\zeta$) of the decay path of log$\sqrt{EM}$ vs. log(T) is also given in the inset for each flare. }
    \label{fig:TvsEM_slope}
    
\end{figure}

Considering pure cooling during the decay phase of the flare, \cite{1991A&A...241..197S} derived single coronal loop length using thermodynamic cooling time scales. Further, \cite{1993A&A...267..586S} explained the sustained heating during the decay of spatially resolved solar flare by introducing the slope in the density-temperature diagram. In the process of explaining the slower decay, \cite{1997A&A...325..782R} included the effect of significant heating during flare decay using a time-dependent hydrodynamic loop model and added a correction factor $F(\zeta)$ to the derived half loop length (L$_d$) as shown in the following equation. 
\begin{equation}
{\rm L_d = 2.7\times10^3\frac{\tau_{d}T_{max}^{1/2}}{F(\zeta)}  ~~cm~~
~~ for ~~~0.35 < \zeta \leq 1.6}
\end{equation}

\begin{equation}
   {\rm  with  \enspace T_{max} = 0.13 T_{0}^{1.16},
    \quad F(\zeta) = \frac{0.51}{\zeta-0.35}+1.36}
    \label{eq:loop}
\end{equation}
Here, $\zeta$ is the slope of the log($\sqrt(EM)$) versus log(T) diagram (equivalent to density -- temperature diagram) during the decay phase of the flare. This diagram is shown in Figure \ref{fig:TvsEM_slope} for the flares F7, F10, F13, and F20, along with the best fit straight line. The $T_{0}$ is the maximum best-fit temperature derived from the spectral fitting of the data in units of K. The F($\zeta$) in equation \ref{eq:loop} is valid for the observation from the EPIC instruments. A similar F($\zeta$) relation was adopted for observations from the RGS instrument (private communication with Reale F.). The slope (${\rm \zeta}$) of the log($\sqrt{EM}$) vs log(T) curve is found to be 0.5$\pm0.1$, 0.9$\pm0.1$, $1.4\pm0.2$, and $0.21\pm0.05$ for flares F7, F10, F13, and F20, respectively. The low values of ${\rm \zeta}$ indicate the presence of sustained heating during the decay of these flares. For the flare F20, the value of {$\rm \zeta$} is out of the range of the model. Therefore, loop length could not be derived from the decay method for F20. For the other flares where the decay phase was observed, we could not use this method as another flare emerged during the decay phase of the previous flare. This situation prevented us from tracing the density-temperature path, which is crucial for determining the loop length. The derived loop lengths for these flares are also given in Table \ref{tab:final_parameters}. These values of the loop lengths are well within 1$\sigma$ with that derived from the flare rise method. 

\subsection{Loop Parameters}
\label{sec:flare_para}
We have calculated the flaring loop volume (V) using the approach used in the past in the absence of direct measurement \cite[see e.g.][]{2000A&A...356..627M,2008MNRAS.387.1627P} by using equation $V = 2 \pi \beta^2 L^3  $ cm$^{3}$, where $\beta$ is the ratio of loop radius to the half loop length. From the solar case, we assume the values of $\beta$ between 0.1 -- 0.3 \citep{1980ApJ...238..343G}. For $\beta$ = 0.3, the V is estimated to be in the range of $3 \times10^{29}$ -- $5 \times 10^{31}$ \vol and found to be maximum for flare F15 and minimum for flare F9.
Using the values of V, $T_{max}$, and peak emission measure (EM), we have derived the plasma density ($n_{e}$), pressure (p) at the loop apex, and minimum magnetic field (B$_{min}$) required to confine the plasma inside the coronal loop. The total plasma density is a sum of electron and hydrogen ion densities (n$_H$). The n$_H$ is found to be 0.8 times n$_e$\footnote{As n$_{He}$ = 0.1 n$_H$, hence n$_H$/n$_e$ = n$_H$/(n$_H$+2n$_{He}$) = 1/1.2 = 0.833}. Thus the values of $n_{e}$, p, V and B$_{min}$ are estimated as:
\begin{equation}
 {\rm n_{e}=\sqrt{\frac{EM}{0.8V}} ~~~ cm^{-3}}; ~~~ {\rm p = 1.8 n_{e}kT_{max} ~ dyne~cm^{-2}};~~~ {\rm B_{min}=\sqrt{8\pi p} ~~G}
\end{equation}

All the estimated values of V, $n_{e}$, p, and B$_{min}$ are given in Table \ref{tab:final_parameters}.
The $n_{e}$ was estimated in the range of 1 to $9\times 10^{11}$ \density, whereas the p was estimated to be in the range of 0.2 to $2\times 10^{4}$ \pressure for all the flares analysed here. The $B_{min}$ was within the range of 200 to 700 Gauss. Among the flares studied, the parameters p and B reached their highest values for flare F1, while the lowest values were observed for flare F5. In a scenario where $\beta$ equals 0.1, these parameters would undergo transformations such that the  V would become $\frac{V}{9}$, $n_e$ would be  $3n_e$,  p would increase to $3p$, and $B_{min}$ would change to $1.73B_{min}$.

The estimated loop lengths were found to be much smaller than the pressure scale height\footnote{$h_p=kT_{max}/ \mu m_{H} g$, where k is Boltzmann's Constant, T$_{max}$ is the maximum temperature, $\mu$ is mean molecular weight, $m_{H}$ is the mass of a hydrogen atom and g is surface gravity of the star.} of the flaring plasma of AB Dor. Therefore, one can assume that the flaring loops are close to a steady-state condition. Thus, the RTV scaling law \citep[][]{1978ApJ...220..643R} can be applied for the estimation of the physical parameters of flaring plasma \cite[see also][]{2008ApJ...672..659A}. 

The heating rate per unit volume (HR$_V$) at the flare peak can be estimated by the following RTV relationship:
\begin{equation}
{\rm HR_{V} \sim 10^{-6} T_{max}^{7/2} L^{-2}}
\end{equation}
Assuming the constant heating during the rise and decay phases of the flare, the total heating rate (E$_{HR}$) can be estimated as {\rm $\left (E_{HR}\sim HR_{V} \times V\right)$}. The total energy corresponding to the heating rate is calculated as {\rm $E_{H,Total}= E_{HR}\times (\tau_{r}+\tau_{d})$}. The estimated values E$_{HR}$ and $E_{H,Total}$ are found to be in the range of 1 $\times$ $10^{31}$ -- 4 $\times$ $10^{33}$ \lum, and 3 $\times$ $10^{34}$ -- 1 $\times$ $10^{37}$ erg, respectively. These parameters were found to be maximum for the strongest flare F15 observed in the sample. 
Under the assumption that the energy released during a flare is of magnetic origin, we have calculated the total non-potential magnetic field ($B_{Total}$) within the active region of the star that corresponds to the release of flare energy. The $B_{Total}$ is found to be in the range of 0.4 to 3.4 kG, as determined by the equation {\rm $E_{H,Total} = (B_{Total}^{2}-B_{min}^2)\times V/8\pi$}. For a comprehensive overview of all estimated loop parameters for AB Dor, please refer to Table \ref{tab:final_parameters}.

\begin{table*}
    \centering
    \caption{Loop parameters.}
    \renewcommand{\arraystretch}{1.2}
    \begin{tabular}{lccccccc}
 \hline
Flare ($\rightarrow$)                   & F1                    & F2                 & F3                 & F4                  & F5                 & F6                  & F7       \\
Parameters ($\downarrow$)               &                       &                    &                    &                     &                      &                     &           \\
\hline
\lxf ($10^{31}$ \lum)                   &  0.982$\pm$0.004      &  1.986$\pm$0.004   &  1.094$\pm$0.003   &  0.890$\pm$0.003    & 0.941$\pm$0.003     & 1.000$\pm$0.004     &  1.268$\pm$0.002  \\
$E_{X,Total}$ ($10^{34}$ erg)           &   3.20$\pm$0.08       &  8.2$\pm$0.4       &  8.2$\pm$0.2       &   4.1$\pm$0.3       & 4.4$\pm$0.3         & 1.17$\pm$0.06       &  7.0$\pm$0.2  \\
$t_{M}$ (ks)                            &  1.2                  & 1.35               & 2.1                &  1.45               & 2.25                & 0.9                 &  3.05 \\
T$_0$ (MK)                              &  44$\pm$2             & 35$\pm$1           & 33$\pm$2           & 34$\pm$1            & 31$\pm$2            &  42$\pm$3           & 36$\pm$1 \\  
T$_M$ (MK)                              &  43$\pm$2             & 30$\pm$1           & 27$\pm$1           & 25$\pm$1            & 25$\pm$2            &  35$\pm$2          & 32$\pm$1  \\  
$L_{r}$ ($10^{10}$ cm)                  &  0.8$\pm$0.1          & 1.0$\pm$0.1        & 1.8$\pm$0.2        & 1.5$\pm$0.2         & 1.8$\pm$0.4         & 0.8$\pm$0.2         &  2.2$\pm$0.2 \\
$L_{d}$ ($10^{10}$ cm)                  &  ...                  & ...                & ...                & ...                 & ...                 &  ...                &  1.6$\pm$0.7 \\
V ($10^{29}$ \vol)                      & 3$\pm$1 & 6$\pm$2 & 30$\pm$13 & 21$\pm$8 & 33$\pm$22 & 3$\pm$2 & 57$\pm$16 \\
$n_{e}$ ($10^{11}$ \density)            & 8$\pm$1 & 5.1$\pm$0.8 & 1.8$\pm$0.4 & 1.8$\pm$0.4 & 1.1$\pm$0.4 & 6$\pm$2 & 1.8$\pm$0.2 \\
p  ($10^4$ \pressure)                   & 2.0$\pm$0.4 & 0.9$\pm$0.1 & 0.32$\pm$0.07 & 0.32$\pm$0.06 & 0.17$\pm$0.06 & 1.4$\pm$0.5 & 0.33$\pm$0.05 \\
$B_{min}$ (G)                           & 703$\pm$69 & 481$\pm$37 & 282$\pm$31 & 285$\pm$28 & 209$\pm$36 & 584$\pm$112 & 288$\pm$21 \\
$B_{Total}$ (kG)                         & 3.4$\pm$0.6 & 1.9$\pm$0.2 & 1.3$\pm$0.2 & 1.3$\pm$0.2 & 0.9$\pm$0.2 & 1.9$\pm$0.5 & 1.1$\pm$0.1 \\
E$_{HR}$ ($10^{31}$ \lum)               & 4$\pm$1 & 1.8$\pm$0.3 & 2.7$\pm$0.7 & 2.8$\pm$0.5 & 2.1$\pm$0.8 & 3$\pm$1 & 4.5$\pm$0.7 \\
$E_{H,Total}$ ($10^{34}$ erg)           & 13$\pm$3 & 8$\pm$1 & 20$\pm$5 & 13$\pm$3 & 10$\pm$4 & 4$\pm$1 & 25$\pm$4 \\
M$_{CME}$ (10$^{18}$ g)                 &   $\sim$2.52          &   $\sim$1.98       &    $\sim$2.07      &   $\sim$1.21        & $\sim$0.97          &  $\sim$0.82        & $\sim$2.57    \\
\hline
Flare ($\rightarrow$)                   & F8                    & F9                 & F10                & F13                 & F15                                      & F20                 & - \\
\hline 
\lxf ($10^{31}$ \lum)                   &  1.750 $\pm$0.005     &  1.743$\pm$0.006   &  1.048$\pm$0.006  &  >0.9$^{a}$               &  >42.4$^{b}$                                   &  3.62 $\pm$ 0.01    & - \\
$E_{X,Total}$ ($10^{34}$ erg)           &   2.0$\pm$0.3         &   12$\pm$2         &  1.5$\pm$0.1       &  >1.4$^{a}$           &  >115$^{b}$                                &  19.4$\pm$0.3       & - \\
t$_{M}$ (ks)                            &  3.55                 & 1.3                & 1.675              & ...                 & 1.95                             &  2.85              & - \\
T$_0$ (MK)                              & 34$\pm$1              & 31$\pm$1           & 41$\pm$4           & 34$\pm$1            & 89$\pm$4                                   & 31$\pm$4           & -\\
T$_M$ (MK)                              & 30$\pm$1              & 29$\pm$1           & 38$\pm$2           & 34$\pm$1            & 56$\pm$2                                   & 24$\pm$1           & - \\
$L_{r}$ ($10^{10}$ cm)                  &  2.6$\pm$0.3          & 0.8$\pm$0.1        & 1.2$\pm$0.3        &  ...                & 4.5$\pm$0.5                               &  2.5$\pm$0.7       & - \\
$L_{d}$ ($10^{10}$ cm)                  &   ...                 &  ...               & 1.2$\pm$0.2        & 1.2$\pm$0.1         & ...                                       &   ...              & - \\
V  ($10^{29}$ \vol)                     & 104$\pm$35 & 2.9$\pm$0.9 & 9$\pm$7 & 10$\pm$2 & 515$\pm$172 & 88$\pm$74 & - \\
$n_{e}$ ($10^{11}$ \density)            & 1.5$\pm$0.2 & 9$\pm$1 & 5$\pm$2 & 4.5$\pm$0.6 & 3.4$\pm$0.6 & 4$\pm$2 & - \\
p  ($10^4$ \pressure)                   & 0.26$\pm$0.05 & 1.5$\pm$0.2 & 1.0$\pm$0.4 & 0.8$\pm$0.1 & 1.9$\pm$0.3 & 0.7$\pm$0.3 & - \\
$B_{min}$ (G)                           & 255$\pm$22 & 617$\pm$48 & 504$\pm$99 & 443$\pm$29 & 682$\pm$59 & 415$\pm$91 & - \\
$B_{Total}$ (kG)                        &  0.4$\pm$0.1 & 2.5$\pm$0.4 & 1.4$\pm$0.4 & 1.0$\pm$0.1 & 2.3$\pm$0.3 & 0.8$\pm$0.3 & - \\
E$_{HR}$ ($10^{31}$ \lum)               & 4.7$\pm$0.8 & 1.0$\pm$0.2 & 4$\pm$2 & 2.0$\pm$0.3 & 379$\pm$73 & 3$\pm$2 & - \\
$E_{H,Total}$ ($10^{34}$ erg)           & 5$\pm$1 & 7$\pm$2 & 6$\pm$3 & 3.0$\pm$0.5 & 1027$\pm$200 & 16$\pm$8 & - \\
M$_{CME}$ (10$^{18}$ g)                 & $\sim$1.10            & $\sim$3.51         &   $\sim$1.39       &   $\sim$1.33        &   $\sim$15.53                           &   7.17               & - \\
 \hline                                                       
    \end{tabular}  
    ~~~\\
    $^a$ Parameters derived using only the decay phase of the flare.\\
    $^b$ Parameters derived using only the rise phase of the flare.\\
    \label{tab:final_parameters}                              
    \end{table*}

\section{Discussion}
\label{sec:discussion}

The quiescent level always exists in the stars even during the flare, so, confining quiescent emission becomes much more important in studies of flares.
We found that the quiescent corona of AB Dor can be well described by three temperature plasma models. In the present study, the average values of T$_{QA}$ and EM$_{QA}$ are found to be 0.94 keV and $4.6\times10^{52}$ \density from {\sc apec} model, and 1.1 keV and $3.1\times10^{52}$ \density from {\sc vapec} model, respectively. These values of T$_{QA}$ and EM$_{QA}$ are consistent with those from the earlier studies of AB Dor \citep[e.g.][]{2001A&A...365L.336G,2003A&A...408.1087S}.
The quiescent state luminosity, L$_{XQ}$ of AB Dor is not found to be constant for different observations analysed here. It was found to be highest during P8 with a value of $1.4\times10^{30}$, \lum whereas, during P4, the value of L$_{XQ}$ of $0.9\times10^{30}$ \lum is found to be lowest.

We have performed a comprehensive study of the 13 strongest X-ray superflares out of a total of 21 observed flares on AB Dor using a large set of data sets from 2000 to 2019 with the XMM-Newton satellite. The time duration of these 21 flares ranges from $\sim$0.7 to $\sim$5.8 hrs. The e-folding rise and decay times of these flares are found to be in the range of 0.27 -- 4.9 ks and 0.7 -- 6.7 ks, respectively, which shows the rapid rise and slower decay pattern of the flares. This kind of trend has been found in many solar and stellar flares \citep[][]{1988A&A...205..181V,2008MNRAS.387.1627P,2012MNRAS.419.1219P,2021MNRAS.505L..79Y}. 
Similar to the flares observed in the past, the F/Q ratios of the flares observed are in the range of 2 -- 4 \citep[e.g.][]{2001A&A...365L.336G,2013A&A...559A.119L}. In the present study, flares F15 and F20 have the highest values of F/Q of 34 and 11, respectively. 
Earlier two strongest X-ray flares were observed in 1997 by BeppoSAX with F/Q of $\sim$100 \citep{2000A&A...356..627M}. These two X-ray flares from BeppoSAX appear to be the strongest flares observed thus far and in our present data the flares F15 and F20, appear to be the next two strongest flares observed in AB Dor thus far.

The detailed TRS analysis shows the variation in spectral parameters during the flaring events. The peak flare temperature was found to be in the range of 31 to 89 MK. Flares F5, F9, and F20 had the lowest peak temperatures, while flare F15 had the highest, and were $\sim$3 to 8 times higher than the quiescent temperature. Whereas the peak emission measure was found to be in the range of 3.2 to $491 \times 10^{52}$ \density in which the maximum value is found for flare F15. Based on the results of TRS, it has been observed that the flare temperature peaks during the rising phase while the emission measure peaks during the peak phase of the flare. This type of delay between peak temperature and peak emission measure has been also found in many solar and stellar flares \citep[][]{1993A&A...267..586S,1989A&A...213..245V,2002A&A...392..585S}. The reason behind this delay is the possible magnetic reconnection process, which leads to the particle acceleration and chromospheric evaporation process \citep{2007A&A...471..271R}. 
The temporal variation in flare temperature of independent single flares and multiple overlapped flares is found to be different. In the case of independent single flares, the temperature rises during the flare's rising phase and decreases during the decay phase due to conduction and radiative cooling. However, in multiple overlapped flares, the temperature remains either constant or increases during the decay phase, as the decay phase of one flare is overlapped with the rising phase of the next flare.

We have also traced the global coronal abundances during the flaring events and found them to be varying as the flare evolves. It is found to be peaked around the peak phase of the flare. We found a 1.1 to 2.7-fold increase in abundances from the quiescent state value ($\sim$0.2 Z$_\odot$) with a maximum increase in abundances for the strongest flare F15. The enhancement in the coronal abundance during the flares could be related to an increase in density within the flaring loop of constant volume which is indicative of a large amount of evaporation of chromospheric material inside the loop due to intense heating during the flares.

The elemental coronal abundances add extra information to the physical nature of the corona of AB Dor. The abundance of the majority of elements was found to be underabundant during the quiescent with respect to the solar photospheric values. However, the abundance of Ne in the quiescent state was found to be more than the solar photospheric value. As the solar photospheric Ne abundances are much uncertain due to the lack of photospheric absorption lines, so, Ne is usually quoted with reference to some other high FIP elements like oxygen, whose photospheric values are well constrained. Here, by assuming the photospheric values of AB Dor similar to the solar photospheric values \citep[][]{2018ApJ...862...66W}, we noticed during the flares (Ne/O)$_*$ ratio ranges from 0.34 to 0.52, whereas the average quiescent value is 0.44 with a minimum of 0.39 and a maximum of 0.48, which is similar to the values reported by \cite{2005Natur.436..525D} for the active stars. Although the FIP effect observed with this assumption was challenged in some cases \citep[studied in detail by][]{2004A&A...416..281S}.
In contrast to solar corona where low FIP elements show enhanced abundances with respect to high FIP elements \citep[][]{1995AdSpR..15g...3V,2000PhyS...61..222F,2015LRSP...12....2L}, the corona of AB Dor shows inverse FIP effect. Such inverse FIP effect is found in active M dwarfs and active binaries \citep[][]{2008A&A...491..859L} and also has been reported for AB Dor by many authors in the past \citep[][]{2000A&A...356..627M,2001A&A...365L.336G,2013A&A...560A..69L}. An inverse-FIP effect near sunspots during flares has recently been detected for the first time \citep{2015ApJ...808L...7D}, which supports that the highly active region shows the I-FIP phenomenon. The Solar-like FIP effects at older aged stars are noticed whereas an inverse-FIP or no FIP effect has been found for the younger (<300 Myr) and most active stars \citep{2005ApJ...622..653T}. Further, it has been reported that there is a strong relationship between spectral class and FIP bias, with M dwarfs having an inverse-FIP effect that decreases to zero at a mid-K spectral type and subsequently drifts toward a solar-like FIP effect for early G dwarfs \citep[][]{2010ApJ...717.1279W,2012ApJ...753...76W}. As AB Dor is a young and highly active K dwarf, therefore, the I-FIP effect is expected. The Fe/O ratio can be used as a proxy to the extent of FIP bias in coronal abundances \cite{2010ApJ...717.1279W}. We found the (Fe/O)$_*$ abundance during the quiescent state ranges from 0.02 -- 0.03. The smaller value for the Fe/O abundance is indicative of a stronger I-FIP effect. The quiescent corona of AB Dor showed the I-FIP effect at nearly the same level for the time span of 19 years, which is longer than the photospheric activity cycle of AB Dor \citep[see][]{2001A&A...365L.336G}. It appears as if the fractionation process that is causing FIP bias in AB Dor is independent of the magnetic activity cycle of AB Dor. Furthermore, the I-FIP effect seems to remain the same or get weaker during the flaring epochs with (Fe/O)$_*$ abundance in the range of 0.02 -- 0.04. For the strongest flare F15 in the current sample, the (Fe/O)$_*$ ratio is found to be a maximum of 0.04, indicating a weaker inverse FIP effect, whereas the overall abundances of each element increase during the flares due to the filling of dense plasma from chromospheric footpoints. %The weaker inverse FIP effect may be due to a larger heating rate in comparison to magnetic energy so the fractionation process weakens. 
One of the possibilities is that the weakened inverse FIP effect is a result of a higher heating rate compared to magnetic energy, which, in turn, weakens the fractionation process.
Although the precise physical mechanisms for the FIP effect are still not fully understood, it is suspected that the drivers of this effect are the Alfven wave heating of the corona and the associated ponderomotive force \citep[][]{2004ApJ...614.1063L,2021ApJ...909...17L}.

%%%%%%%%%%%Loop parameters%%%%%%%%%%%%%

Using the hydrodynamic loop model, we have derived the semi-loop length of the flares and found it to be in the range of 0.8 -- 4.5 $\times$ $10^{10}$ cm. The estimated loop length is also found to be similar to the earlier observed flares on AB Dor \citep[][]{2001A&A...365L..36M,2001A&A...365L.336G, 2007MNRAS.377.1488H,2013A&A...560A..69L}. The highest loop length in AB Dor was found to be $\sim$ 8.8 $\times$ $10^{10}$ cm in November 2009 observations \citep[][]{2013A&A...560A..69L}. In the case of the Sun, the typical loop length is found to be of the order of 10$^9$ - 10$^{10}$ cm \citep[][]{2009phsu.book.....M}. The loop heights (2L/$\pi$) corresponding to semi-loop lengths of all these flaring events observed so far are 8 to 43\% of the radius of AB Dor. 
The total energy released by the flares was estimated to be in the range of 3 $\times$ 10$^{34}$ - 1 $\times$ 10$^{37}$ erg, which is very large in comparison to the total energy of the strongest flares observed on the Sun \citep[$\sim$10$^{32}$ erg][]{2012ApJ...759...71E, 2020ApJ...891..138Z}. Also, the total magnetic field of the loop is found to be around 0.4 - 3.4 kG, which is the typical magnetic field observed in AB Dor \citep[][]{1997MNRAS.291....1D, 1999MNRAS.305L..35J}.

If we closely inspect the PN light curve of the post-flare phase of set S2, we notice a continuous dimming as shown by \citet{2021NatAs...5..697V}. A rotational modulation appears in most X-ray light curves of AB Dor, possibly due to the stellar surface's eclipsing of the coronal active regions (Singh et al., private communication). 
After removing the effect of rotational modulation from set S2 and following the definition of dimming as given in \citet{2021NatAs...5..697V}, we didn't see any strong signature of such dimming in the light curve (See Figure \ref{fig:rot_mod}).
An empirical relationship between the stellar flare energy in X-rays and its associated CME mass is estimated as $M_{CME} (g) = 10^{-1.5\pm0.5} E_{G}^{0.59\pm0.02}$, where $E _{G}$ is the X-ray energy in GOES (1 -- 8 \AA) energy band \citep{2012ApJ...760....9A,2013ApJ...764..170D}. 
The derived X-ray flux is converted into GOES flux using WEBPIMMS for the derived flare temperatures of AB Dor as described in Section \ref{sec:TRS_flare}. The estimated values of $M_{CME}$ for AB Dor are found to be in the range $10^{18 - 19}$ g and found to be maximum for the flare F15. These values of CMEs are 10 to 100 times more than the most massive solar CME \citep{2009IAUS..257..233Y} and similar to other stellar CMEs \citep{2022MNRAS.509.3247K, 2022NatAs...6..241N, 1994A&A...285..489G}.

\section{Summary and Conclusions}
\label{sec:summary}
We have analysed quiescent and flaring X-ray emission of the active fast rotating star, AB Dor. In most of the observations, we observed rotational modulation in the quiescent state light curves. The quiescent state of AB Dor consists of three temperature plasma with an average value of the temperature, emission measure, and abundances of 0.94 keV, $4.6 \times 10^{52}$ \density, and 0.2 Z$_\odot$, respectively. The quiescent state luminosity of AB Dor was not found to be constant over the 19 years of observations supporting the presence of long-term variations. A total of 21 flares are detected from six observations of AB Dor with a flare-to-quiescent state count rate ratio of 2 -- 4 for the majority of flares. 
The most powerful flares observed in AB Dor are identified as F15 and F20, with flare-to-quiescent state count rate ratios of 34 and 11, respectively. The flare F15, which occurred in 2016, emerges as the third most powerful flare following the two strongest flares documented during the BeppoSAX observations in 1997.  The most intense flare, F15, exhibits a peak temperature of 89 MK, unlike the other flares, which all have peak temperatures below 50 MK. 
 In most flares, we observe an increase in abundance and density, suggesting chromospheric evaporation. The elemental abundances exhibit an inverse FIP bias in both quiescent and flaring conditions. The heights of the loops in these flares do not extend beyond 50 per cent of the stellar radius. Additionally, the erupted mass of CMEs appears to be 10 to 100 times higher than the most massive solar CME.

\section*{Acknowledgements}
This work is based on the observations obtained from the XMM–Newton, an ESA science mission with instruments and contributions funded by ESA Member States and NASA. SD acknowledges CSIR for providing the grant for her research work. AKS acknowledges the support of the ISRO project for his scientific research. We thank the reviewer, Prof. Antonio Maggio, for his useful comments and suggestions.
%%%%%%%%%%%%%%%%%%%%%%%%%%%%%%%%%%%%%%%%%%%%%%%%%%
\section*{Data Availability}
The data used in the paper is available through the XMM-Newton archive.



\bibliographystyle{mnras}
\bibliography{ms.bbl} % if your bibtex file is called example.bib

%%%%%%%%%%%%%%%%% APPENDICES %%%%%%%%%%%%%%%%%%%%%

%\newpage
\onecolumn
%\lipsum[1,10]
\appendix
\section{Spectral evolution of flares}
%\begin{center}
\begin{longtable}{cccccccccc}
\caption{Best fit spectral parameters of each temporal segment of the flares F1-F10, F13, F15, and F20. Here, FS represents flare segments, and ST and ET refer to the start and end times, respectively of each flare segment relative to the start time of the corresponding observation.
} \label{tab:all_flares_2+1apec} \\
%\begin{tabular}
\hline
     Set &Flare&FS  & ST:ET (ks)                  & kT$_{3}$                & EM$_{3}$             & Z                                     & \lxf                   & $\chi_\nu^2 $ (dof)\\
          &     &    &                             &(keV)                   &(10$^{52}$ \density) & (Z$_\odot$)               & ($10^{30}$ \lum)&                    \\
 \endfirsthead

\multicolumn{10}{l}%
{{\bfseries \tablename\ \thetable{} -- continued from previous page}} \\
\hline 
      Set &Flare&FS  & ST:ET (ks)                  & kT$_{3}$               & EM$_{3}$            & Z                                & \lxf                   & $\chi_\nu^2 $ (dof)\\
          &     &    &                             &(keV)                   &(10$^{52}$ \density) & (Z$_\odot$)                & ($10^{30}$ \lum)&                    \\ \hline

\endhead
\hline \multicolumn{10}{r}{{Continued on next page}} \\ 
\endfoot    

\endlastfoot
          
         \hline     
         S1& F1 & R  &   4.8 : 5.7       &  3.8$_{-0.2}^{+0.2}$    & 5.4$_{-0.2}^{+0.2}$ & 0.193$_{-0.005}^{+0.004}$ & 1.71$_{-0.01}^{+0.01}$ & 1.1 (352)\\
           &    & P  &  5.7 : 6.3        &  3.7$_{-0.1}^{+0.1}$    & 15.7$_{-0.3}^{+0.3}$ & 0.255$_{-0.007}^{+0.007}$ & 3.34$_{-0.02}^{+0.02}$ & 1.54 (426)\\
           &    & D1 &  6.3 : 6.9        &  2.49$_{-0.1}^{+0.09}$  & 11.0$_{-0.3}^{+0.3}$ & 0.281$_{-0.007}^{+0.007}$ & 2.5$_{-0.02}^{+0.02}$ & 1.33 (350)\\
           &    & D2 &  6.9 : 7.4        &  2.8$_{-0.1}^{+0.1}$    & 9.6$_{-0.3}^{+0.3}$ & 0.243$_{-0.007}^{+0.007}$ & 2.27$_{-0.02}^{+0.02}$ & 1.03 (317)\\
           & F2 & R  &  7.4 : 8.2        &  3.0$_{-0.1}^{+0.1}$    & 11.1$_{-0.3}^{+0.3}$ & 0.225$_{-0.007}^{+0.007}$ & 2.45$_{-0.02}^{+0.02}$ & 1.05 (360)\\
           &    & P  &  8.2 : 9.3        &  2.56$_{-0.07}^{+0.09}$ & 11.9$_{-0.3}^{+0.2}$ & 0.248$_{-0.006}^{+0.006}$ & 2.52$_{-0.01}^{+0.01}$ & 1.0 (432)\\
           &    & D1 &  9.3 : 10.1       &  2.37$_{-0.07}^{+0.07}$ & 10.2$_{-0.3}^{+0.3}$ & 0.246$_{-0.006}^{+0.006}$ & 2.28$_{-0.01}^{+0.01}$ & 0.98 (374)\\
           &    & D2 &  10.1 : 11.0      &  2.35$_{-0.08}^{+0.08}$ & 7.9$_{-0.2}^{+0.2}$ & 0.232$_{-0.005}^{+0.005}$ & 1.97$_{-0.01}^{+0.01}$ & 1.16 (365)\\
           &    & D3 &  11.0 : 12.0      &  2.38$_{-0.09}^{+0.1}$  & 6.2$_{-0.2}^{+0.2}$ & 0.206$_{-0.005}^{+0.005}$ & 1.7$_{-0.01}^{+0.01}$ & 1.29 (350)\\
           &    & D4 &  12.0 : 13.2      &  2.6$_{-0.1}^{+0.1}$    & 4.7$_{-0.2}^{+0.2}$ & 0.209$_{-0.004}^{+0.004}$ & 1.56$_{-0.01}^{+0.01}$ & 1.08 (359)\\
           &    & D5 &  13.2 : 14.4      &  2.3$_{-0.1}^{+0.1}$    & 4.3$_{-0.2}^{+0.2}$ & 0.185$_{-0.004}^{+0.004}$ & 1.42$_{-0.01}^{+0.01}$ & 1.05 (343)\\
           &    & D6 &  14.4 : 15.7      &  2.5$_{-0.1}^{+0.2}$    & 3.9$_{-0.2}^{+0.2}$ & 0.181$_{-0.004}^{+0.004}$ & 1.38$_{-0.01}^{+0.01}$ & 1.1 (349)\\
           &    & D7 &  15.7 : 17.0      &  2.6$_{-0.2}^{+0.2}$    & 2.8$_{-0.2}^{+0.2}$ & 0.174$_{-0.004}^{+0.004}$ & 1.24$_{-0.01}^{+0.01}$ & 1.04 (330)\\
           &    & D8 &  17.0 : 18.3      &  2.6$_{-0.2}^{+0.2}$    & 2.2$_{-0.1}^{+0.1}$ & 0.174$_{-0.003}^{+0.003}$ & 1.17$_{-0.01}^{+0.01}$ & 1.03 (336)\\
           &    & D9 &  18.3 : 20.0      &  2.5$_{-0.2}^{+0.2}$    & 1.9$_{-0.2}^{+0.1}$ & 0.169$_{-0.003}^{+0.003}$ & 1.12$_{-0.01}^{+0.01}$ & 1.07 (328)\\
           &    & D10&  20.0 : 21.4      &  3.0$_{-0.3}^{+0.4}$    & 1.3$_{-0.1}^{+0.1}$ & 0.167$_{-0.003}^{+0.003}$ & 1.05$_{-0.01}^{+0.01}$ & 1.19 (310)\\
           & F3 & R  &  24.3 : 25.6      &  2.9$_{-0.1}^{+0.2}$    & 4.2$_{-0.2}^{+0.2}$ & 0.173$_{-0.004}^{+0.004}$ & 1.41$_{-0.01}^{+0.01}$ & 1.12 (358)\\
           &    & P  &  25.6 : 27.2      &  2.33$_{-0.08}^{+0.06}$ & 8.2$_{-0.2}^{+0.2}$ & 0.216$_{-0.004}^{+0.004}$ & 1.94$_{-0.01}^{+0.01}$ & 1.35 (446)\\
           &    & D1 &  27.2 : 28.8      &  2.10$_{-0.06}^{+0.06}$ & 6.3$_{-0.2}^{+0.2}$ & 0.219$_{-0.004}^{+0.004}$ & 1.71$_{-0.01}^{+0.01}$ & 1.26 (414)\\
           &    & D2 &  28.8 : 30.6      &  2.14$_{-0.06}^{+0.06}$ & 5.7$_{-0.2}^{+0.2}$ & 0.204$_{-0.004}^{+0.004}$ & 1.61$_{-0.01}^{+0.01}$ & 1.14 (418)\\
           &    & D3 &  30.6 : 32.6      &  2.16$_{-0.06}^{+0.07}$ & 5.5$_{-0.1}^{+0.1}$ & 0.187$_{-0.003}^{+0.003}$ & 1.53$_{-0.01}^{+0.01}$ & 1.2 (428)\\
           &    & D4 &  32.6 : 34.3      &  2.13$_{-0.07}^{+0.08}$ & 4.7$_{-0.2}^{+0.2}$ & 0.182$_{-0.004}^{+0.004}$ & 1.43$_{-0.01}^{+0.01}$ & 1.0 (386)\\
           &    & D5 &  34.3 : 36.4      &  2.02$_{-0.07}^{+0.08}$ & 3.8$_{-0.1}^{+0.1}$ & 0.176$_{-0.003}^{+0.003}$ & 1.31$_{-0.01}^{+0.01}$ & 1.02 (391)\\
         S2& F4 & R  &    4.1 : 5.1      & 3.0$_{-0.1}^{+0.1}$ & 3.6$_{-0.2}^{+0.2}$ & 0.202$_{-0.005}^{+0.005}$ & 1.27$_{-0.01}^{+0.01}$ & 1.28 (299)\\
           &    & P  &     5.1 : 6.0     & 2.15$_{-0.08}^{+0.1}$ & 5.4$_{-0.2}^{+0.2}$ & 0.228$_{-0.005}^{+0.006}$ & 1.54$_{-0.01}^{+0.01}$ & 1.12 (334)\\
           &    & D1 &     6.0 : 6.7     & 1.89$_{-0.1}^{+0.1}$ & 3.7$_{-0.2}^{+0.2}$ & 0.224$_{-0.006}^{+0.006}$ & 1.33$_{-0.01}^{+0.01}$ & 1.15 (281)\\
           &    & D2 &     6.7 : 7.5     & 1.93$_{-0.1}^{+0.1}$ & 3.2$_{-0.2}^{+0.2}$ & 0.217$_{-0.005}^{+0.005}$ & 1.26$_{-0.01}^{+0.01}$ & 1.09 (287)\\
           &    & D3 &     7.5 : 8.3     & 1.86$_{-0.1}^{+0.1}$ & 2.9$_{-0.2}^{+0.2}$ & 0.208$_{-0.005}^{+0.005}$ & 1.21$_{-0.01}^{+0.01}$ & 1.16 (281)\\
           &    & D4 &     8.3 : 9.1     & 2.15$_{-0.2}^{+0.2}$ & 2.6$_{-0.2}^{+0.2}$ & 0.203$_{-0.005}^{+0.005}$ & 1.18$_{-0.01}^{+0.01}$ & 1.26 (279)\\
           &    & D5 &     9.1 : 10.5    & 1.73$_{-0.1}^{+0.1}$ & 2.2$_{-0.2}^{+0.2}$ & 0.2$_{-0.005}^{+0.005}$ & 1.11$_{-0.01}^{+0.01}$ & 1.11 (276)\\
           & F5 & R1 &     10.5 : 11.7   & 1.96$_{-0.1}^{+0.2}$ & 2.2$_{-0.2}^{+0.2}$ & 0.208$_{-0.005}^{+0.005}$ & 1.14$_{-0.01}^{+0.01}$ & 1.03 (262)\\
           &    & R2 &     11.7 : 12.4   & 2.66$_{-0.2}^{+0.2}$ & 2.8$_{-0.2}^{+0.2}$ & 0.22$_{-0.005}^{+0.005}$ & 1.28$_{-0.01}^{+0.01}$ & 1.08 (265)\\
           &    & P  &     12.4 : 13.1   & 2.14$_{-0.1}^{+0.2}$ & 3.2$_{-0.2}^{+0.2}$ & 0.231$_{-0.006}^{+0.006}$ & 1.32$_{-0.01}^{+0.01}$ & 1.11 (264)\\
           &    & D1 &     13.1 : 13.9   & 1.93$_{-0.1}^{+0.1}$ & 3.1$_{-0.2}^{+0.2}$ & 0.211$_{-0.006}^{+0.006}$ & 1.24$_{-0.01}^{+0.01}$ & 1.06 (275)\\
           &    & D2 &     13.9 : 14.7   & 1.78$_{-0.1}^{+0.1}$ & 2.8$_{-0.2}^{+0.2}$ & 0.209$_{-0.006}^{+0.006}$ & 1.2$_{-0.01}^{+0.01}$ & 1.09 (267)\\
           &    & D3 &     14.7 : 15.5   & 1.93$_{-0.1}^{+0.1}$ & 2.1$_{-0.2}^{+0.2}$ & 0.21$_{-0.005}^{+0.005}$ & 1.14$_{-0.01}^{+0.01}$ & 1.08 (259)\\
           &    & D4 &     15.5 : 16.3   & 1.86$_{-0.2}^{+0.2}$ & 1.7$_{-0.2}^{+0.2}$ & 0.209$_{-0.005}^{+0.005}$ & 1.09$_{-0.01}^{+0.01}$ & 1.25 (251)\\
           &    & D5 &     16.3 : 17.3   & 1.82$_{-0.2}^{+0.2}$ & 0.9$_{-0.1}^{+0.2}$ & 0.205$_{-0.005}^{+0.004}$ & 1.0$_{-0.01}^{+0.01}$ & 1.28 (259)\\
           & F6 & R  &     29.1 : 29.8   & 3.6$_{-0.23}^{+0.26}$ & 4.7$_{-0.2}^{+0.2}$ & 0.198$_{-0.005}^{+0.005}$ & 1.51$_{-0.01}^{+0.01}$ & 1.31 (301)\\
           &    & P  &     29.8 : 30.2   & 2.99$_{-0.15}^{+0.17}$ & 8.6$_{-0.3}^{+0.3}$ & 0.256$_{-0.007}^{+0.008}$ & 2.1$_{-0.02}^{+0.02}$ & 1.17 (310)\\  
           &    & D1 &     30.2 : 30.6   & 2.99$_{-0.17}^{+0.2}$ & 6.6$_{-0.3}^{+0.3}$ & 0.247$_{-0.007}^{+0.007}$ & 1.84$_{-0.02}^{+0.02}$ & 1.11 (287)\\
           &    & D2 &     30.6 : 31.1   & 3.21$_{-0.19}^{+0.22}$ & 5.6$_{-0.2}^{+0.2}$ & 0.216$_{-0.006}^{+0.006}$ & 1.65$_{-0.01}^{+0.01}$ & 0.97 (289)\\
           &    & D3 &     31.1 : 31.6   & 3.27$_{-0.22}^{+0.25}$ & 4.6$_{-0.2}^{+0.2}$ & 0.201$_{-0.006}^{+0.006}$ & 1.49$_{-0.01}^{+0.01}$ & 0.97 (277)\\
           &    & D4 &     31.6 : 32.6   & 2.92$_{-0.19}^{+0.2}$ & 4.5$_{-0.2}^{+0.2}$ & 0.181$_{-0.006}^{+0.006}$ & 1.4$_{-0.01}^{+0.01}$ & 1.05 (268)\\
           & F7 & R1 &     32.6 : 34.1   & 3.08$_{-0.1}^{+0.1}$ & 6.8$_{-0.2}^{+0.2}$ & 0.194$_{-0.004}^{+0.004}$ & 1.74$_{-0.01}^{+0.01}$ & 1.13 (434)\\
           &    & R2 &     34.1 : 35.1   & 2.98$_{-0.08}^{+0.08}$ & 11.6$_{-0.2}^{+0.2}$ & 0.202$_{-0.005}^{+0.005}$ & 2.35$_{-0.01}^{+0.01}$ & 1.03 (453)\\
           &    & P  &     35.1 : 36.2   & 2.76$_{-0.06}^{+0.06}$ & 14.4$_{-0.2}^{+0.2}$ & 0.218$_{-0.005}^{+0.005}$ & 2.65$_{-0.01}^{+0.01}$ & 1.21 (488)\\
           &    & D1 &     36.2 : 37.6   & 2.51$_{-0.05}^{+0.06}$ & 11.2$_{-0.2}^{+0.2}$ & 0.224$_{-0.005}^{+0.005}$ & 2.25$_{-0.01}^{+0.01}$ & 1.26 (482)\\
           &    & D2 &     37.6 : 39.2   & 2.48$_{-0.06}^{+0.06}$ & 8.6$_{-0.2}^{+0.2}$ & 0.215$_{-0.004}^{+0.004}$ & 1.92$_{-0.01}^{+0.01}$ & 1.01 (460)\\
           &    & D3 &     39.2 : 42.0   & 2.31$_{-0.06}^{+0.06}$ & 8.0$_{-0.2}^{+0.2}$ & 0.197$_{-0.004}^{+0.004}$ & 1.77$_{-0.01}^{+0.01}$ & 1.1 (457)\\
           & F8 & R1 &     42.0 : 42.9   & 2.26$_{-0.09}^{+0.1}$ & 6.9$_{-0.2}^{+0.2}$ & 0.226$_{-0.006}^{+0.006}$ & 1.73$_{-0.01}^{+0.01}$ & 1.02 (327)\\
           &    & R2 &     42.9 : 43.6   & 2.6$_{-0.1}^{+0.1}$ & 8.6$_{-0.3}^{+0.3}$ & 0.249$_{-0.007}^{+0.007}$ & 2.02$_{-0.01}^{+0.01}$ & 1.1 (342)\\
           &    & R3 &     43.6 : 44.2   & 2.5$_{-0.1}^{+0.1}$ & 10.0$_{-0.3}^{+0.3}$ & 0.244$_{-0.007}^{+0.007}$ & 2.18$_{-0.02}^{+0.02}$ & 1.08 (332)\\
           &    & R4 &     44.2 : 44.8   & 2.8$_{-0.1}^{+0.1}$ & 12.5$_{-0.3}^{+0.3}$ & 0.219$_{-0.007}^{+0.007}$ & 2.48$_{-0.02}^{+0.02}$ & 1.11 (360)\\
           &    & R5 &     44.8 : 45.3   & 3.0$_{-0.1}^{+0.1}$ & 16.88$_{-0.4}^{+0.4}$ & 0.215$_{-0.008}^{+0.008}$ & 3.04$_{-0.02}^{+0.02}$ & 1.11 (376)\\
           &    & P  &     45.3 : 45.8   & 2.56$_{-0.07}^{+0.1}$ & 17.6$_{-0.5}^{+0.4}$ & 0.225$_{-0.009}^{+0.01}$ & 3.01$_{-0.02}^{+0.02}$ & 1.17 (369)\\
           &    & D  &     45.8 : 46.3   & 2.42$_{-0.07}^{+0.09}$ & 17.4$_{-0.5}^{+0.4}$ & 0.257$_{-0.009}^{+0.01}$ & 3.04$_{-0.02}^{+0.02}$ & 0.95 (368)\\
           & F9 & R1 &     46.3 : 46.7   & 2.46$_{-0.09}^{+0.08}$ & 17.4$_{-0.4}^{+0.5}$ & 0.29$_{-0.01}^{+0.01}$ & 3.17$_{-0.02}^{+0.02}$ & 1.0 (339)\\
           &    & R2 &     46.7 : 47.1   & 2.47$_{-0.08}^{+0.07}$ & 20.3$_{-0.5}^{+0.5}$ & 0.27$_{-0.01}^{+0.01}$ & 3.46$_{-0.03}^{+0.03}$ & 1.03 (356)\\
           &    & R3 &     47.1 : 47.4   & 2.71$_{-0.08}^{+0.09}$ & 20.1$_{-0.5}^{+0.5}$ & 0.32$_{-0.01}^{+0.01}$ & 3.72$_{-0.03}^{+0.03}$ & 1.24 (371)\\
           &    & P  &     47.4 : 47.8   & 2.54$_{-0.07}^{+0.07}$ & 20.5$_{-0.5}^{+0.5}$ & 0.32$_{-0.01}^{+0.01}$ & 3.69$_{-0.03}^{+0.03}$ & 0.92 (366)\\
           &    & D  &     47.8 : 48.2   & 2.22$_{-0.06}^{+0.06}$ & 19.3$_{-0.5}^{+0.5}$ & 0.31$_{-0.01}^{+0.01}$ & 3.39$_{-0.02}^{+0.02}$ & 1.04 (355)\\
         S3& F10& R  &   39.04 : 40.49   & 3.5$_{-0.3}^{+0.3}$ & 6.8$_{-0.4}^{+0.4}$ & 0.25$_{-0.01}^{+0.01}$ & 1.87$_{-0.03}^{+0.02}$ & 0.88 (208)\\
           &    & P  &   40.49 : 40.94   & 3.2$_{-0.2}^{+0.2}$ & 16.7$_{-0.7}^{+0.7}$ & 0.25$_{-0.02}^{+0.02}$ & 3.15$_{-0.04}^{+0.04}$ & 1.02 (230)\\
           &    & D1 &   40.94 : 41.49   & 2.5$_{-0.2}^{+0.2}$ & 11.2$_{-0.5}^{+0.5}$ & 0.28$_{-0.01}^{+0.01}$ & 2.34$_{-0.03}^{+0.03}$ & 1.14 (252)\\
           &    & D2 &   41.49 : 42.19   & 2.1$_{-0.1}^{+0.1}$ & 7.6$_{-0.4}^{+0.4}$ & 0.23$_{-0.01}^{+0.01}$ & 1.73$_{-0.02}^{+0.02}$ & 1.09 (228)\\
           &    & D3 &   42.19 : 43.04   & 1.8$_{-0.1}^{+0.2}$ & 4.8$_{-0.4}^{+0.4}$ & 0.23$_{-0.01}^{+0.01}$ & 1.39$_{-0.02}^{+0.02}$ & 0.94 (216)\\
         S4& F13& P  &     37.63 : 38.23 & 2.9$_{-0.1}^{+0.1}$   & 14.6$_{-0.5}^{+0.5}$ & 0.33$_{-0.02}^{+0.02}$ & 2.68$_{-0.03}^{+0.03}$ & 0.97 (268)\\
           &    & D1 &     38.23 : 38.63 & 2.6$_{-0.1}^{+0.1}$ & 10.5$_{-0.3}^{+0.4}$ & 0.33$_{-0.01}^{+0.01}$ & 2.11$_{-0.02}^{+0.02}$ & 1.15 (272)\\
           &    & D2 &     38.63 : 39.13 & 2.07$_{-0.1}^{+0.08}$ & 9.1$_{-0.3}^{+0.4}$ & 0.28$_{-0.01}^{+0.01}$ & 1.72$_{-0.02}^{+0.02}$ & 1.08 (270)\\
           &    & D3 &     39.13 :39.63  & 1.67$_{-0.08}^{+0.07}$ & 6.7$_{-0.3}^{+0.3}$ & 0.28$_{-0.01}^{+0.01}$ & 1.4$_{-0.02}^{+0.02}$ & 1.24 (240)\\
           &    & D4 &     39.63 :40.23  & 1.57$_{-0.06}^{+0.05}$ & 5.6$_{-0.2}^{+0.3}$ & 0.29$_{-0.01}^{+0.01}$ & 1.31$_{-0.01}^{+0.01}$ & 1.14 (244)\\
         S5& F15& R1 &   8.19 : 9.39     & 5.7$_{-0.3}^{+0.3}$ & 30.0$_{-0.6}^{+0.6}$ & 0.29$_{-0.01}^{+0.01}$ & 6.40$_{-0.05}^{+0.05}$ & 1.03 (411)\\
           &    & R2 &   9.39 : 9.59     & 7.4$_{-0.3}^{+0.3}$ & 199$_{-3}^{+3}$ & 0.50$_{-0.04}^{+0.04}$ & 38.9$_{-0.3}^{+0.3}$ & 1.09 (561)\\
           &    & R3 &   9.59 : 9.69     & 7.7$_{-0.3}^{+0.4}$ & 315$_{-5}^{+5}$ & 0.53$_{-0.05}^{+0.06}$ & 61.5$_{-0.4}^{+0.4}$ & 1.14 (557)\\
           &    & R4 &   9.69 : 9.79     & 6.7$_{-0.2}^{+0.3}$ & 405$_{-7}^{+7}$ & 0.41$_{-0.05}^{+0.06}$ & 73.7$_{-0.6}^{+0.6}$ & 1.21 (472)\\
           &    & R5 &   9.79 : 9.99     & 5.8$_{-0.2}^{+0.2}$ & 461$_{-7}^{+7}$ & 0.40$_{-0.04}^{+0.04}$ & 81.0$_{-0.6}^{+0.6}$ & 1.06 (532)\\
           &    & R6 &   9.99 : 10.09    & 5.3$_{-0.2}^{+0.2}$ & 450$_{-9}^{+9}$ & 0.54$_{-0.06}^{+0.06}$ & 80.6$_{-0.7}^{+0.7}$ & 1.12 (432)\\
           &    & P  &   10.09 : 10.19   & 4.8$_{-0.1}^{+0.2}$ & 491$_{-9}^{+9}$ & 0.42$_{-0.05}^{+0.05}$ & 81.9$_{-0.7}^{+0.7}$ & 1.15 (451)\\
         S6&*F20& R  &   68.0 : 70.2     & 2.7$_{-0.2}^{+0.3}$    & 47.2$_{-0.5}^{+0.6}$ & **              &  4.80$_{-0.04}^{+0.04}$    & 1.28 (657)\\
           &    & P  &   70.2 : 71.5     & 2.07$_{-0.08}^{+0.05}$ & 132$_{-1}^{+1}$      & **              &  12.19$_{-0.09}^{+0.08}$   & 1.35 (906)\\
           &    & D1 &   71.5 : 73.3     & 1.97$_{-0.03}^{+0.03}$ & 92.4$_{-0.7}^{+0.7}$ & **              &  8.84$_{-0.06}^{+0.06}$    & 1.76 (925)\\
           &    & D2 &   73.3 : 76.2     & 1.79$_{-0.02}^{+0.02}$ & 55.1$_{-0.5}^{+0.5}$ & **              &  5.57$_{-0.04}^{+0.04}$    & 1.95 (962)\\
           &    & D3 &   76.2 : 81.1     & 1.68$_{-0.03}^{+0.03}$ & 27.8$_{-0.3}^{+0.3}$ & **              &  3.16$_{-0.02}^{+0.02}$    & 1.90 (947)\\
           &    & D4 &   81.1 : 90.3     & 1.65$_{-0.06}^{+0.03}$ & 10.7$_{-0.1}^{+0.1}$  & **             &  1.62$_{-0.01}^{+0.01}$    & 1.62 (948)\\
         \hline
%\end{tabular}

\multicolumn{10}{l}
{\footnotesize * Spectral fitting was carried out using RGS spectra.}\\
\multicolumn{10}{l}
{\footnotesize ** The abundances were fixed to the quiescent state values of segment P11.}
\end{longtable}
%\end{center}

%%%%%%%%%%%%%%%%%%%%%%%%%%%%%%%%%%%%%%%%%%%%%%%%%%

\begin{landscape}
\begin{table*}
{\scriptsize
\setlength{\tabcolsep}{1.2pt}
    \renewcommand{\arraystretch}{0.9}
    \centering
    \caption{Best fit spectral parameters as obtained from the spectral fitting of RGS spectra. All the parameters are quoted with a 68 per cent confidence interval.}     \label{tab:3vapec_qui_all}
    \hskip-2.0cm\begin{tabular}{ccccccccccccccccccccc}
         \hline\hline
         Parameters ($\rightarrow$) & $T_{1}$ & $T_{2}$ & $T_{3}$ & $EM_{1}$ & $EM_{2}$ & $EM_{3}$ & C & N & O & Ne & Mg & Si & S & Ar & Fe & \lxf & $\chi_\nu^2$(dof) \\
         Segments ($\downarrow$) & (keV) & (keV) & (keV) & ($10^{52}$ \density) & ($10^{52}$ \density) & ($10^{52}$ \density) & & & & & & & & & & (10$^{30}$ \lum)& \\
         \hline
         S2-P3 & 0.28$_{-0.01}^{+0.01}$ & 0.76$_{-0.02}^{+0.03}$ & 1.7$_{-0.2}^{+0.2}$ & 1.9$_{-0.2}^{+0.2}$ & 4.2$_{-0.4}^{+0.4}$ & 3.9$_{-0.3}^{+0.3}$ & 0.7$_{-0.1}^{+0.1}$ & 0.7$_{-0.1}^{+0.1}$ & 0.37$_{-0.04}^{+0.04}$ & 1.2$_{-0.1}^{+0.1}$ & 0.22$_{-0.08}^{+0.08}$ & 0.3$_{-0.1}^{+0.1}$ & 0.2$_{-0.1}^{+0.1}$ & 0.9$_{-0.3}^{+0.3}$ & 0.17$_{-0.02}^{+0.02}$ & 1.07$_{-0.01}^{+0.01}$ & 1.31 (434) \\
         \\
         F4 & - & - & 1.66$_{-0.08}^{+0.1}$ & - & - & 4.1$_{-0.2}^{+0.2}$ & 0.87$_{-0.09}^{+0.09}$ & 0.73$_{-0.09}^{+0.09}$ & 0.39$_{-0.01}^{+0.01}$ & 1.39$_{-0.08}^{+0.08}$ & 0.32$_{-0.07}^{+0.07}$ & 0.3$_{-0.1}^{+0.1}$ & 0.5$_{-0.1}^{+0.1}$ & 0.7$_{-0.2}^{+0.2}$ & 0.182$_{-0.006}^{+0.006}$ & 1.14$_{-0.01}^{+0.01}$ & 1.34 (583) \\
         \\
         F5 & - & - & 1.8$_{-0.1}^{+0.2}$ & - & - & 2.7$_{-0.1}^{+0.1}$ & 0.81$_{-0.07}^{+0.07}$ & 0.49$_{-0.06}^{+0.06}$ & 0.39$_{-0.01}^{+0.01}$ & 1.29$_{-0.06}^{+0.06}$ & 0.29$_{-0.06}^{+0.06}$ & 0.27$_{-0.07}^{+0.07}$ & 0.34$_{-0.07}^{+0.07}$ & 0.8$_{-0.1}^{+0.1}$ & 0.175$_{-0.004}^{+0.004}$ & 0.973$_{-0.007}^{+0.007}$ & 1.60 (811) \\
         \\
         P4 & 0.30$_{-0.02}^{+0.01}$ & 0.73$_{-0.03}^{+0.02}$ & 1.8$_{-0.3}^{+0.4}$ & 1.6$_{-0.2}^{+0.2}$ & 3.1$_{-0.3}^{+0.4}$ & 2.1$_{-0.2}^{+0.2}$ & 0.9$_{-0.2}^{+0.2}$ & 0.7$_{-0.1}^{+0.1}$ & 0.45$_{-0.04}^{+0.04}$ & 1.5$_{-0.1}^{+0.2}$ & 0.37$_{-0.08}^{+0.09}$ & 0.6$_{-0.1}^{+0.1}$ & 0.6$_{-0.1}^{+0.2}$ & 1.1$_{-0.3}^{+0.3}$ & 0.19$_{-0.02}^{+0.02}$ & 0.811$_{-0.008}^{+0.008}$ & 1.31 (471) \\
         \\
         F6 & - & - & 1.9$_{-0.1}^{+0.2}$ & - & - & 6.3$_{-0.3}^{+0.3}$ & 0.9$_{-0.1}^{+0.1}$ & 0.6$_{-0.1}^{+0.1}$ & 0.35$_{-0.02}^{+0.02}$ & 1.2$_{-0.1}^{+0.1}$ & 0.6$_{-0.1}^{+0.1}$ & 0.8$_{-0.2}^{+0.2}$ & 0.6$_{-0.2}^{+0.2}$ & 0.9$_{-0.3}^{+0.3}$ & 0.199$_{-0.008}^{+0.008}$ & 1.38$_{-0.02}^{+0.02}$ & 1.14 (348)\\
         \\
         F7 & - & - & 1.94$_{-0.07}^{+0.09}$ & - & - & 10.2$_{-0.2}^{+0.2}$ & 0.75$_{-0.07}^{+0.07}$ & 0.64$_{-0.07}^{+0.07}$ & 0.35$_{-0.01}^{+0.01}$ & 1.16$_{-0.06}^{+0.06}$ & 0.24$_{-0.06}^{+0.06}$ & 0.45$_{-0.09}^{+0.09}$ & 0.11$_{-0.08}^{+0.08}$ & 1.1$_{-0.1}^{+0.1}$ & 0.180$_{-0.005}^{+0.005}$ & 1.65$_{-0.01}^{+0.01}$ & 1.19 (1041) \\
         \\
         F8 & - & - & 1.92$_{-0.06}^{+0.09}$ & - & - & 15.0$_{-0.3}^{+0.3}$ & 0.89$_{-0.09}^{+0.09}$ & 0.7$_{-0.1}^{+0.1}$ & 0.41$_{-0.01}^{+0.01}$ & 1.30$_{-0.08}^{+0.08}$ & 0.30$_{-0.07}^{+0.07}$ & 0.6$_{-0.1}^{+0.1}$ & 0.4$_{-0.1}^{+0.1}$ & 1.1$_{-0.2}^{+0.2}$ & 0.220$_{-0.007}^{+0.007}$ & 2.25$_{-0.01}^{+0.01}$ & 1.20 (971) \\
         \\
         F9 & - & - & 1.87$_{-0.09}^{+0.1}$ & - & - & 19.9$_{-0.6}^{+0.6}$ & 1.1$_{-0.2}^{+0.2}$ & 0.6$_{-0.2}^{+0.2}$ & 0.41$_{-0.02}^{+0.02}$ & 1.2$_{-0.1}^{+0.1}$ & 0.3$_{-0.1}^{+0.1}$ & 0.5$_{-0.2}^{+0.2}$ & 0.3$_{-0.2}^{+0.2}$ & 1.5$_{-0.4}^{+0.4}$ & 0.27$_{-0.01}^{+0.01}$ & 2.81$_{-0.03}^{+0.03}$ & 1.19 (425) \\
         \\
         S3-P5 & 0.294$_{-0.004}^{+0.004}$ & 0.75$_{-0.01}^{+0.01}$ & 1.61$_{-0.03}^{+0.05}$ & 1.94$_{-0.08}^{+0.09}$ & 3.6$_{-0.1}^{+0.1}$ & 4.4$_{-0.1}^{+0.1}$ & 0.77$_{-0.05}^{+0.05}$ & 0.59$_{-0.04}^{+0.05}$ & 0.39$_{-0.01}^{+0.01}$ & 1.14$_{-0.04}^{+0.04}$ & 0.34$_{-0.03}^{+0.03}$ & 0.39$_{-0.04}^{+0.04}$ & 0.37$_{-0.05}^{+0.05}$ & 0.70$_{-0.09}^{+0.09}$ & 0.205$_{-0.007}^{+0.008}$ & 1.105$_{-0.004}^{+0.004}$ & 1.46 (1923) \\
         \\
         F10 & - & - & 1.7$_{-0.1}^{+0.2}$ & - & - & 9.0$_{-0.4}^{+0.4}$ & 0.8$_{-0.2}^{+0.2}$ & 0.3$_{-0.1}^{+0.1}$ & 0.44$_{-0.03}^{+0.03}$ & 1.2$_{-0.1}^{+0.1}$ & 0.3$_{-0.1}^{+0.1}$ & 0.3$_{-0.2}^{+0.2}$ & 1.0$_{-0.3}^{+0.3}$ & 1.5$_{-0.5}^{+0.5}$ & 0.161$_{-0.008}^{+0.008}$ & 1.58$_{-0.02}^{+0.02}$ & 1.18 (363) \\
         \\
         S4-F11 & - & - & 1.75$_{-0.09}^{+0.07}$ & - & - & 8.7$_{-0.2}^{+0.2}$ & 0.9$_{-0.1}^{+0.1}$ & 0.7$_{-0.1}^{+0.1}$ & 0.48$_{-0.02}^{+0.02}$ & 1.20$_{-0.09}^{+0.08}$ & 0.29$_{-0.07}^{+0.07}$ & 0.4$_{-0.1}^{+0.1}$ & 0.6$_{-0.2}^{+0.2}$ & 1.4$_{-0.3}^{+0.3}$ & 0.206$_{-0.006}^{+0.006}$ & 1.51$_{-0.01}^{+0.01}$ & 1.12 (707) \\
         \\
         P6 & 0.28$_{-0.01}^{+0.01}$ & 0.70$_{-0.01}^{+0.01}$ & 1.68$_{-0.06}^{+0.09}$ & 1.5$_{-0.1}^{+0.1}$ & 3.8$_{-0.2}^{+0.2}$ & 5.1$_{-0.2}^{+0.2}$ & 0.86$_{-0.09}^{+0.09}$ & 0.59$_{-0.07}^{+0.07}$ & 0.44$_{-0.02}^{+0.02}$ & 1.20$_{-0.07}^{+0.07}$ & 0.36$_{-0.05}^{+0.05}$ & 0.32$_{-0.07}^{+0.07}$ & 0.5$_{-0.1}^{+0.1}$ & 1.2$_{-0.2}^{+0.2}$ & 0.17$_{-0.01}^{+0.01}$ & 1.134$_{-0.007}^{+0.007}$ & 1.14 (1111) \\
         \\ 
         F12 & - & - & 1.69$_{-0.05}^{+0.09}$ & - & - & 8.9$_{-0.2}^{+0.2}$ & 0.6$_{-0.1}^{+0.1}$ & 0.9$_{-0.1}^{+0.1}$ & 0.47$_{-0.02}^{+0.02}$ & 1.36$_{-0.08}^{+0.08}$ & 0.33$_{-0.07}^{+0.07}$ & 0.4$_{-0.1}^{+0.1}$ & 0.5$_{-0.1}^{+0.1}$ & 1.1$_{-0.3}^{+0.3}$ & 0.209$_{-0.006}^{+0.006}$ & 1.54$_{-0.01}^{+0.01}$ & 1.09 (717) \\
         \\
         F13 & - & - & 2.1$_{-0.2}^{+0.2}$ & - & - & 11.1$_{-0.4}^{+0.4}$ & 1.1$_{-0.2}^{+0.2}$ & 0.6$_{-0.2}^{+0.2}$ & 0.49$_{-0.03}^{+0.03}$ & 1.4$_{-0.1}^{+0.1}$ & 0.3$_{-0.1}^{+0.1}$ & 0.50$_{-0.2}^{+0.2}$ & 0.7$_{-0.3}^{+0.3}$ & 0.6$_{-0.5}^{+0.5}$ & 0.200$_{-0.009}^{+0.009}$ & 1.75$_{-0.02}^{+0.02}$ & 1.12 (405) \\
         \\
         P7 & 0.29$_{-0.01}^{+0.01}$ & 0.70$_{-0.02}^{+0.02}$ & 1.59$_{-0.09}^{+0.1}$ & 1.3$_{-0.2}^{+0.2}$ & 3.8$_{-0.3}^{+0.3}$ & 4.6$_{-0.3}^{+0.3}$ & 1.1$_{-0.1}^{+0.2}$ & 0.6$_{-0.1}^{+0.1}$ & 0.47$_{-0.04}^{+0.04}$ & 1.4$_{-0.1}^{+0.1}$ & 0.24$_{-0.07}^{+0.07}$ & 0.6$_{-0.1}^{+0.1}$ & 0.4$_{-0.2}^{+0.2}$ & 1.3$_{-0.3}^{+0.4}$ & 0.17$_{-0.02}^{+0.02}$ & 1.10$_{-0.01}^{+0.01}$ & 1.24 (510) \\
         \\
         F14 & - & - & 1.69$_{-0.05}^{+0.09}$ & - & - & 6.3$_{-0.2}^{+0.2}$ & 1.0$_{-0.1}^{+0.1}$ & 0.78$_{-0.09}^{+0.09}$ & 0.49$_{-0.01}^{+0.01}$ & 1.42$_{-0.07}^{+0.07}$ & 0.34$_{-0.06}^{+0.06}$ & 0.37$_{-0.08}^{+0.08}$ & 1.0$_{-0.1}^{+0.1}$ & 1.2$_{-0.2}^{+0.2}$ & 0.196$_{-0.005}^{+0.005}$ & 1.324$_{-0.009}^{+0.009}$ & 1.13 (935) \\
         \\
         S5-P8 & 0.297$_{-0.01}^{+0.009}$ & 0.76$_{-0.02}^{+0.03}$ & 1.6$_{-0.1}^{+0.1}$ & 2.1$_{-0.2}^{+0.2}$ & 3.5$_{-0.2}^{+0.3}$ & 4.0$_{-0.2}^{+0.2}$ & 0.9$_{-0.1}^{+0.1}$ & 0.9$_{-0.1}^{+0.1}$ & 0.44$_{-0.03}^{+0.04}$ & 1.3$_{-0.1}^{+0.1}$ & 0.29$_{-0.07}^{+0.07}$ & 0.5$_{-0.1}^{+0.1}$ & 0.8$_{-0.2}^{+0.2}$ & 0.6$_{-0.2}^{+0.2}$ & 0.22$_{-0.02}^{+0.02}$ & 1.134$_{-0.007}^{+0.007}$ & 1.22 (596) \\
         \\
         F15 & - & - & 4.94$_{-0.07}^{+0.07}$ & - & - & 120.6$_{-0.7}^{+0.7}$ & 2.4$_{-0.1}^{+0.1}$ & 1.6$_{-0.1}^{+0.1}$ & 1.12$_{-0.02}^{+0.02}$ & 3.00$_{-0.08}^{+0.08}$ & 0.72$_{-0.09}^{+0.09}$ & 0.7$_{-0.1}^{+0.1}$ & 2.7$_{-0.2}^{+0.2}$ & 3.4$_{-0.4}^{+0.4}$ & 0.90$_{-0.01}^{+0.01}$ & 15.05$_{-0.03}^{+0.03}$ & 1.59 (3111) \\
         \\
         F16 & - & - & 1.91$_{-0.08}^{+0.1}$ & - & - & 12.5$_{-0.5}^{+0.6}$ & 1.7$_{-0.3}^{+0.4}$ & 1.3$_{-0.3}^{+0.3}$ & 0.96$_{-0.05}^{+0.05}$ & 2.6$_{-0.2}^{+0.2}$ & 0.6$_{-0.1}^{+0.2}$ & 0.6$_{-0.2}^{+0.2}$ & 0.7$_{-0.4}^{+0.4}$ & 1.5$_{-0.6}^{+0.7}$ & 0.44$_{-0.02}^{+0.02}$ & 2.51$_{-0.03}^{+0.03}$ & 1.13 (3110) \\
         \\
         P9 & 0.294$_{-0.004}^{+0.004}$ & 0.74$_{-0.01}^{+0.01}$ & 1.69$_{-0.03}^{+0.04}$ & 1.53$_{-0.06}^{+0.06}$ & 3.23$_{-0.09}^{+0.09}$ & 4.72$_{-0.08}^{+0.08}$ & 0.86$_{-0.05}^{+0.05}$ & 0.70$_{-0.04}^{+0.04}$ & 0.45$_{-0.01}^{+0.01}$ & 1.38$_{-0.04}^{+0.04}$ & 0.38$_{-0.03}^{+0.03}$ & 0.31$_{-0.04}^{+0.04}$ & 0.49$_{-0.05}^{+0.05}$ & 0.81$_{-0.09}^{+0.09}$ & 0.202$_{-0.006}^{+0.006}$ & 1.10$_{-0.01}^{+0.01}$ & 1.39 (2282) \\
         \\
         S6-F17 & - & - & 1.71$_{-0.08}^{+0.09}$ & - & - & 5.5$_{-0.2}^{+0.2}$ & 1.4$_{-0.2}^{+0.2}$ & 0.8$_{-0.1}^{+0.1}$ & 0.54$_{-0.02}^{+0.02}$ & 1.28$_{-0.09}^{+0.08}$ & 0.40$_{-0.08}^{+0.08}$ & 0.4$_{-0.1}^{+0.1}$ & 0.6$_{-0.2}^{+0.2}$ & 0.8$_{-0.3}^{+0.3}$ & 0.208$_{-0.007}^{+0.007}$ & 1.20$_{-0.01}^{+0.01}$ & 1.19 (531) \\
         \\
         F18 & - & - & 1.88$_{-0.08}^{+0.09}$ & - & - & 6.6$_{-0.2}^{+0.2}$ & 1.3$_{-0.1}^{+0.2}$ & 0.7$_{-0.1}^{+0.1}$ & 0.53$_{-0.02}^{+0.02}$ & 1.25$_{-0.07}^{+0.07}$ & 0.29$_{-0.07}^{+0.07}$ & 0.5$_{-0.1}^{+0.1}$ & 0.4$_{-0.1}^{+0.1}$ & 0.9$_{-0.2}^{+0.2}$ & 0.213$_{-0.006}^{+0.006}$ & 1.29$_{-0.01}^{+0.01}$ & 1.11 (822) \\
         \\
         P11 & 0.302$_{-0.008}^{+0.02}$ & 0.76$_{-0.01}^{+0.03}$ & 2.3$_{-0.2}^{+0.3}$ & 1.5$_{-0.1}^{+0.2}$ & 3.0$_{-0.2}^{+0.2}$ & 3.4$_{-0.2}^{+0.2}$ & 1.1$_{-0.1}^{+0.2}$ & 0.7$_{-0.1}^{+0.1}$ & 0.53$_{-0.03}^{+0.04}$ & 1.6$_{-0.1}^{+0.1}$ & 0.31$_{-0.07}^{+0.07}$ & 0.3$_{-0.1}^{+0.1}$ & 0.6$_{-0.1}^{+0.1}$ & 0.5$_{-0.2}^{+0.2}$ & 0.19$_{-0.01}^{+0.02}$ & 0.962$_{-0.005}^{+0.005}$ & 1.18 (749) \\
         \\
         F19 & - & - & 1.79$_{-0.07}^{+0.1}$ & - & - & 8.8$_{-0.3}^{+0.3}$ & 0.9$_{-0.2}^{+0.2}$ & 0.8$_{-0.2}^{+0.2}$ & 0.46$_{-0.02}^{+0.02}$ & 1.3$_{-0.1}^{+0.1}$ & 0.5$_{-0.1}^{+0.1}$ & 0.6$_{-0.1}^{+0.1}$ & 0.8$_{-0.2}^{+0.2}$ & 0.5$_{-0.3}^{+0.3}$ & 0.250$_{-0.009}^{+0.009}$ & 1.6$_{-0.01}^{+0.01}$ & 0.95 (509)\\
         \\
         F20 & - & - & 2.78$_{-0.05}^{+0.05}$ & - & - & 32.4$_{-0.2}^{+0.2}$ & 1.3$_{-0.1}^{+0.1}$ & 0.93$_{-0.08}^{+0.08}$ & 0.72$_{-0.01}^{+0.01}$ & 2.13$_{-0.06}^{+0.06}$ & 0.50$_{-0.06}^{+0.06}$ & 0.51$_{-0.08}^{+0.08}$ & 1.0$_{-0.1}^{+0.1}$ & 1.4$_{-0.2}^{+0.2}$ & 0.457$_{-0.006}^{+0.006}$ & 4.32$_{-0.01}^{+0.01}$ & 1.42 (2440) \\
         \\
         F21 & - & - & 1.9$_{-0.1}^{+0.1}$ & - & - & 7.0$_{-0.2}^{+0.2}$ & 1.2$_{-0.2}^{+0.2}$ & 0.7$_{-0.1}^{+0.1}$ & 0.61$_{-0.02}^{+0.02}$ & 1.59$_{-0.09}^{+0.09}$ & 0.45$_{-0.08}^{+0.09}$ & 0.6$_{-0.1}^{+0.1}$ & 0.4$_{-0.2}^{+0.2}$ & 0.7$_{-0.3}^{+0.3}$ & 0.251$_{-0.007}^{+0.007}$ & 1.44$_{-0.01}^{+0.01}$ & 1.21 (682) \\
         \\
         \hline
    \end{tabular}

}
\end{table*}
\end{landscape}

% Don't change these lines

\bsp	% typesetting comment

\label{lastpage}
\end{document}